\documentclass[manuscript,screen]{acmart}
\usepackage{algorithmic}
\usepackage{graphicx}
\usepackage{textcomp}
\usepackage{balance}
\usepackage{xcolor}
\usepackage{bm}
\usepackage{amsmath}
\usepackage{caption}
\usepackage{subcaption}
\usepackage{multirow}
\usepackage{adjustbox}
\usepackage{booktabs}
\usepackage{makecell}
\usepackage[]{hyperref}
\usepackage{wrapfig}
\usepackage{xcolor}
\usepackage[ruled,vlined]{algorithm2e}
\usepackage{url}

\AtBeginDocument{%
  \providecommand\BibTeX{{%
    \normalfont B\kern-0.5em{\scshape i\kern-0.25em b}\kern-0.8em\TeX}}}

\setcopyright{acmcopyright}
\copyrightyear{2021}
\acmYear{2021}
\acmDOI{10.1145/1122445.1122456}

\begin{document}

\title{An Adaptive Graph Pre-training Framework for Localized Collaborative Filtering}



\author{Yiqi Wang}
\email{wangy206@msu.edu}
\affiliation{%
  \institution{Michigan State University}
  \country{USA}
}

\author{Chaozhuo Li}
\email{cli@microsoft.com}
\affiliation{%
  \institution{Microsoft Research Asia}
  \country{China}
}

\author{Zheng Liu}
\email{zheng.liu@microsoft.com}
\affiliation{%
  \institution{Microsoft Research Asia}
  \country{China}
}

\author{Mingzheng Li}
\email{mingzhengli@microsoft.com}
\affiliation{%
  \institution{Microsoft}
  \country{China}
}

\author{Jiliang Tang}
\email{tangjili@msu.edu}
\affiliation{%
  \institution{Michigan State University}
  \country{USA}
}

\author{Xing Xie}
\email{xingx@microsoft.com}
\affiliation{%
  \institution{Microsoft Research Asia}
  \country{China}
}

\author{Lei Chen}
\email{leichen@cse.ust.hk}
\affiliation{%
  \institution{Hong Kong University of Science and Technology}
  \country{China}
}

\author{Philip S. Yu}
\email{psyu@uic.edu}
\affiliation{%
  \institution{University of Illinois at Chicago}
  \country{USA}
}

\renewcommand{\shortauthors}{Wang and Li, et al.}

\begin{abstract}
Graph neural networks (GNNs) have been widely applied in the recommendation tasks and have obtained very appealing performance. However, most GNN-based recommendation methods suffer from the problem of data sparsity in practice. Meanwhile, pre-training techniques have achieved great success in mitigating data sparsity in various domains such as natural language processing (NLP) and computer vision (CV). Thus, graph pre-training has the great potential to alleviate data sparsity in GNN-based recommendations. However, pre-training GNNs for recommendations face unique challenges. For example, user-item interaction graphs in different recommendation tasks have distinct sets of users and items, and they often present different properties. Therefore, the successful mechanisms commonly used in NLP and CV to transfer knowledge from pre-training tasks to downstream tasks such as sharing learned embeddings or feature extractors are not directly applicable to existing GNN-based recommendations models.  To tackle these challenges, we delicately design an adaptive graph pre-training framework for localized collaborative filtering (ADAPT). It does not require transferring user/item embeddings, and is able to capture both the common knowledge across different graphs and the uniqueness for each graph. Extensive experimental results have demonstrated the effectiveness and superiority of ADAPT.
\end{abstract}



\maketitle

\section{Introduction}

Recommendation is one of the most ubiquitous and successful applications of artificial intelligence in our daily life. It has been widely adopted in various online services such as target advertising and online shopping. 
Basically, recommendation systems aim at predicting a user's preference based on her historical interactions with different items, and further recommending the user some items that she may have potential interest in~\cite{adomavicius2005toward}. Many existing solutions for recommendations follow the paradigm to first learn a set of latent factors (i.e., embeddings for users and items) and then build an interaction function to make recommendation decisions based on the learned embeddings. Matrix factorization (MF) is one representative method of such techniques. It aims at learning user and item embeddings directly from the user-item interaction matrix and then makes predictions via inner product over the user and item embeddings. In recent years, we have witnessed increasing efforts on incorporating deep neural networks to advance these techniques via refining the user and item embeddings~\cite{he2017neural} and modeling interaction functions~\cite{wang2015collaborative,guo2017deepfm}.

The interactions among users and items in a recommendation task can be naturally denoted as a user-item bipartite graph. From this perspective, the key for a recommendation task is to learn the node representations from the user-item bipartite graph.  Graph neural networks (GNNs), which generalize deep neural networks (DNNs) to graph data, have been theoretically and empirically proved to be very powerful in representation learning for graph data~\cite{wu2020comprehensive,zhang2020deep}.  Therefore, there is increasing attention on adopting GNNs in addressing recommendation tasks. GNNs are able to inherently capture important high-order user-item connectivity in a given user-item interaction bipartite graph, as a consequence, they can boost the recommendation performance. For example, NGCF~\cite{wang2019neural} proposed to propagate embeddings over user-item graph based on the message-passing framework of GNNs and achieved significant performance improvement; and LightGCN~\cite{he2020lightgcn} further refined the design of GNNs for recommendation tasks by eliminating the feature transformation and non-linear activation, and achieved the state of the art performance. 
Despite achieving great success in recommendation tasks, we empirically found that current GNN-based recommendation methods suffer from a common and practical problem: data scarcity\footnote{More details about how the data sparsity affects the performance of existing GNN-based recommendation methods can be found at Section~\ref{sec:experiment}.}. In other words, when the user-item interaction graphs are sparse, the performance of most existing GNN-based recommendations will tend to drop substantially.  Unfortunately, historical user-item interactions in the real-world recommendations are often scarce~\cite{adomavicius2005toward}. 
The challenge of data scarce is also universal in other domains such as NLP~\cite{qiu2020pre} and CV~\cite{he2019rethinking,huh2016makes}, where pre-training techniques have been proposed to alleviate this problem. Typically, a model is first pre-trained on a large dataset with abundant label information (either self-supervised label or supervised label), and then finetuned over the downstream datasets with limited label information. Pre-training has been demonstrated to be effective in alleviating the data scarcity issue of the downstream task by transferring knowledge from the pre-training data. Thus, it is natural to ask: \textit{can we also leverage pre-training techniques to facilitate GNN-based recommendation models?}

Adopting pre-training for existing GNN-based recommendations faces unique challenges. The great success of pre-training in NLP and CV relies on the effective mechanisms that enable knowledge sharing or transferring from the pre-trained tasks to the downstream tasks. In NLP, the vocabulary of words or tokens is shared; therefore, both context-independent embeddings (e.g., word2vec~\cite{mikolov2013efficient}) and context-sensitive embeddings (e.g., GPT~\cite{radford2018improving} and BERT~\cite{devlin2018bert} from pre-training tasks can be transferred. In CV,  low-/mid-level features (such as edges, textures) given by the pre-trained model can be leveraged by the fine-tuned tasks. However, user and item embeddings denote the major parameters of existing GNN-based recommendation models. Moreover, in recommendations, different interaction graphs often do not have the same sets of users and items. Therefore, these uniquenesses determine that the effective mechanisms from NLP and CV are not applicable to existing GNN-based recommendations. In addition, a large amount of data is essential for the effectiveness of pre-training~\cite{zhang2020you} in NLP and CV. For instance, the state-of-the-art pre-trained model in CV is typically pre-trained on tens of millions of images~\cite{deng2009imagenet}, and the pre-training dataset for NLP often consists of more than 1000M words~\cite{devlin2018bert}. Therefore, it is desired to take advantage of many interaction graphs for pre-training. However, achieving this goal is challenging given that different graphs have distinct properties such as the number of items (or users) and graph density. As a consequence, dedicated efforts are required to tackle these unique challenges.

In this work, we propose an~\textbf{Ada}ptive graph \textbf{P}re-training framework for localized collaborative fil\textbf{T}ering (\textbf{ADAPT}) that provides effective solutions to address the aforementioned challenges. It consists of two key components: a meta localized GNN (or meta-LGNN) model and the GNN adaptor. The meta-LGNN model provides a new perspective to build GNN-based recommendations, where it is trained to make recommendation predictions based on the neighboring structures of the target user and item, instead of learning a set of user and item embeddings and an interaction function. The rationality of this design is:  the key collaborative filtering information of a given target user and target item can be encoded by their neighboring structure, which consists of their historical interactions. With meta-LGNN, there is no need to transfer user or item embeddings across different graphs.
To leverage multiple pre-training graphs with distinct properties, we design the GNN adaptor as a strategy to capture their differences. Specifically, given a recommendation task, the GNN adaptor can adapt the meta-LGNN model to a customized GNN model by considering the properties of its corresponding interaction graph. In summary, our contributions can be summarized as follows:
\begin{itemize}
    \item We design a meta-LGNN model which is trained to make recommendation predictions based on the neighboring structure around the target user and item. Compared to the existing GNN-based recommendations which require to learn a set of user and item embeddings, this design paves a way to enable pre-training for GNN-based recommendations from a new perspective.
    \item We propose an adaptive pre-training framework based on meta-LGNN. It allows pre-training with multiple user-item interaction graphs while considering the uniqueness of each graph.  
    \item We have conducted extensive experiments on various datasets, and the empirical results demonstrate the effectiveness and superiority of the proposed framework.
\end{itemize}
The remaining of the paper is organized as follows. In Section 2, we describe the proposed framework in details, including the problem definition, framework overview, the basis model and two key processes. We introduce the experiments to validate the effectiveness in Section 3, including experimental settings, preliminary study, performance comparison, ablation study and further probing. In Section 4, we then review some important related work. We conclude the paper with discussions on future work in Section 5.

\section{The Proposed Framework}

In this section, we first formally define the GNN pre-training problem for GNN-based recommendations. Then we describe an overview of the proposed ADAPT. Next, we introduce the novel GNN  recommendation method for localized collaborative filtering as the basis of the proposed framework. Finally, we detail the pre-training process and the fine-tuning process. 

\subsection{The GNN Pre-Training Problem on Recommendations}

We first briefly describe the problem studied in this paper. It is common that user-item interaction graphs are sparse in the real-world recommendations~\cite{adomavicius2005toward}. It is undoubtedly very challenging to perform GNN-based recommendations on a sparse user-item interaction graph. Meanwhile, there exist a large amount of user-item bipartite graphs from other recommendation tasks, which contain abundant information and knowledge.
A natural idea to mitigate the data sparsity problem is to transfer useful information from existing graphs to facilitate the recommendation task on the target sparse graph. Recent years have witnessed the great success of  the pre-training techniques to transfer knowledge from a large amount of data to help solve the target task in NLP and CV~\cite{radford2018improving,devlin2018bert}. 
Therefore, in this paper, we mainly focus on leveraging the pre-training strategy to address the data sparsity challenge in GNN-based recommendations. 

Next, we formally define the pre-training problem for GNN-based recommendations. We denote the user-item interaction data in a recommendation task as a user-item bipartite graph $G= (U,I,E)$, where $U=\{u_1,u_2,\cdots,u_{|U|}\}$ represents the user set,  $I=\{i_1,i_2,\cdots,i_{|I|}\}$ is the item set, and $E \subset {U \times I}$ indicates interactions between users and items. Each element $e_{jk}$ in $E$ suggests that there exists an interaction between the user $u_j$ and the item $i_k$.  We assume that there are $N$ graphs for pre-training that are denoted $\{G_1,G_2,\cdots,G_N\}$. We further use $G_t= (U_t,I_t,E_t)$ to denote the target graph for the downstream recommendation task.

Generally, a GNN model consists of $L$ layers, where the key component in each layer $l$ is a graph convolution with parameter ${\bm \theta}^l_{conv}$. Thus, we can denote the parameters for a GNN model as $ {\bm \Theta}_{GNN} = \{{\bm \theta}^1_{conv},\cdots,{\bm \theta}^L_{conv}\}$. With above definitions, the goal of the pre-training task is to train a model from graphs $\{G_1,G_2,\cdots,G_N\}$ that can help build a GNN model $ \mathcal{G}(\cdot;{\bm \Theta}_{GNN_{t}})$ for $G_t$ from the downstream recommendation task. 
\begin{figure}[t]
    \centering
    \includegraphics[width=0.8\linewidth]{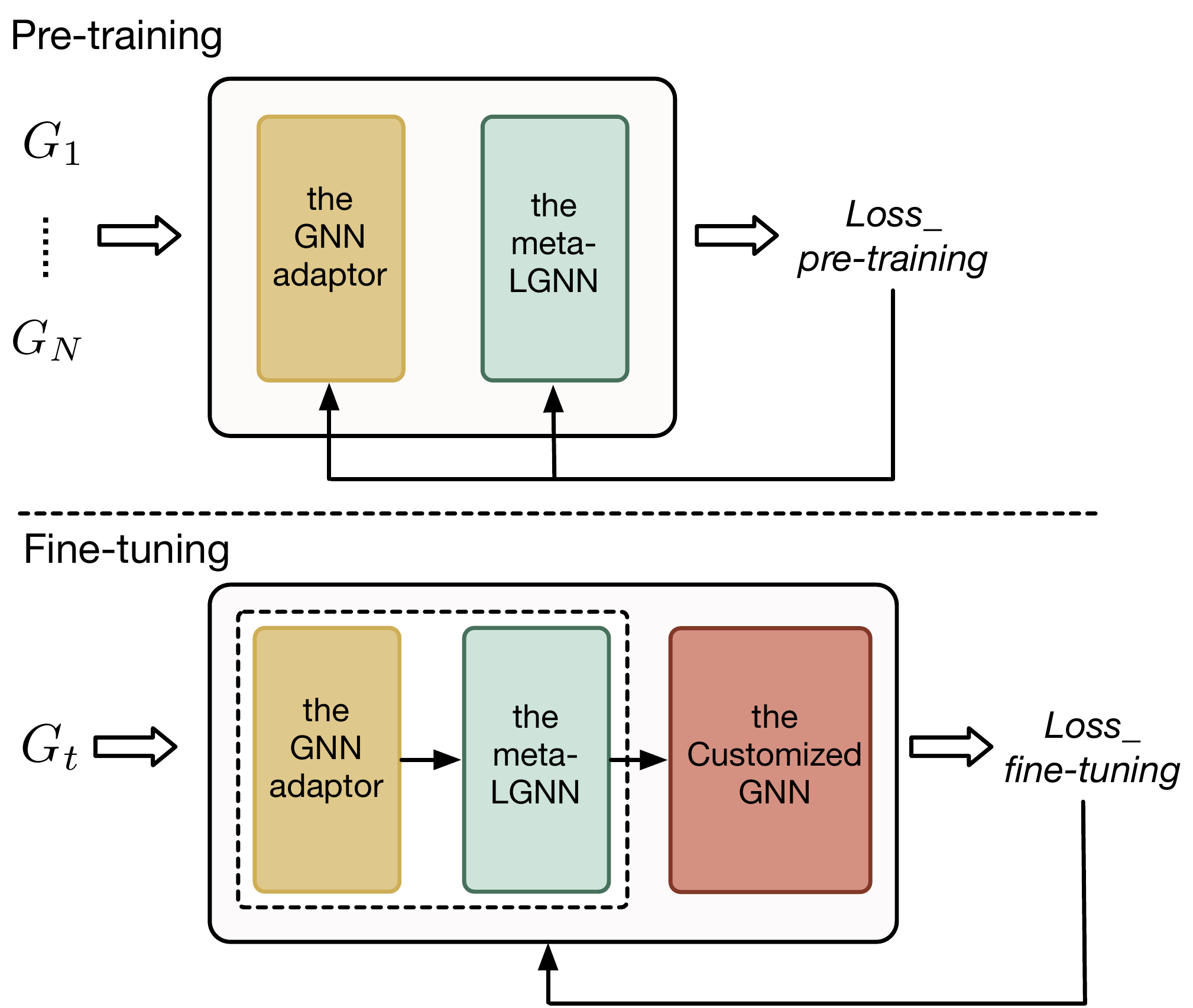}
    \caption{An overview of the proposed framework ADAPT. There are two key components in ADAPT -- a meta-LGNN model and a GNN adaptor. The GNN adaptor is designed to adapt the meta-LGNN model to a customized GNN model for the interaction graph of a given recommendation task by considering its distinct properties. With these two components, ADAPT first trains a GNN adaptor and a meta-LGNN model simultaneously in the pre-training phase. Then, given the target graph from the downstream recommendation task in the fine-tuning phase, we generate a customized GNN model based on the properties of the target graph via the GNN adaptor and the meta-LGNN model.}
    \label{fig:frame}
\end{figure}

\subsection{An Overall Design of ADAPT}

An overview of ADAPT is demonstrated in Figure~\ref{fig:frame}. It consists of two key components -- a meta-LGNN model and a GNN adaptor. One key challenge for pre-training GNNs for recommendations is that user-item interaction graphs in different recommendation tasks have distinct user/item sets; as a result, it is hard to directly transfer user/item embeddings across different recommendation graphs like pre-trained word embeddings in NLP. To solve this challenge, we propose the meta-LGNN model. It provides a new perspective for GNN-based recommendations, where the recommendation prediction is made based on the local structure surrounding a user and an item, rather than their embeddings, via a GNN model. Therefore, it does not need to learn embeddings of users and items, which is naturally more flexible for pre-training, compared to most existing GNN-based recommendation methods. The success behind pre-training is that there exists common knowledge that can be transferred from the pre-training data to the downstream task. On the one hand, there are similar patterns of user-item local structures in different recommendation graphs, which are captured by the proposed meta-LGNN model. However, there could exist different patterns for these graphs since different graphs can present distinct properties. Therefore, ADAPT provides the GNN adaptor that can adapt the meta-LGNN model to a customized GNN model for the interaction graph of a given recommendation task by considering its distinct properties. With these two components, ADAPT first trains a GNN adaptor and a meta-LGNN model simultaneously in the pre-training phase. Then, given the target graph from the downstream recommendation task in the fine-tuning phase, we generate a customized GNN model based on the properties of the target graph via the GNN adaptor and the meta-LGNN model. In the following subsections, we first introduce the proposed GNN recommendation method for localized collaborative filtering in ADAPT. Next, we describe the pre-training process and the fine-tuning process of ADAPT in detail.

\subsection{GNN Recommendation Method for Localized Collaborative Filtering}
Graph Neural Network(GNN) models have been demonstrated effectiveness and superior in facilitating recommendation tasks~\cite{yang2018hop,ying2018graph,wang2019neural,he2020lightgcn}. Most existing GNN-based recommendation methods aim at learning a set of user and item embeddings via a GNN model. Given that user-item interaction graphs from different recommendation tasks have distinct user/item sets,  these GNN-based recommendation methods are impractical to be pre-trained, since user/item embeddings can not be transferred across different graphs. To take advantage of the power of both GNN methods and the pre-training techniques, we propose a new GNN-based localized collaborative filtering (meta-LGNN), which does not require to learn user/item embeddings and makes predictions based on local structures extracted from the interaction graphs. Next, we will first briefly introduce the most common existing GNN-based recommendation method, and then illustrate the meta-LGNN.

We use $\textbf{H} = \{\textbf{h}_{u_1},\cdots,\textbf{h}_{u_{|U|}},\textbf{h}_{i_1},\cdots,\textbf{h}_{i_{|I|}}\}$ to denote the user and item embeddings the model aims to learn.  $ {\bm \Theta}_{GNN} = \{{\bm \theta}^1_{conv},\cdots,{\bm \theta}^L_{conv}\}$ represents the parameters to be learned for a $L$-layer GNN model, where ${\bm \theta}^l_{conv}$ denotes the parameters for the graph convolution in layer $l$. Generally, for a user $u$ in a given user item bipartite graph, the graph convolution first aggregates item embeddings from its neighborhood, and then updates the target user embedding based on its original embeddings and the aggregated embeddings. The update process of item embeddings works similarly as that for a user node. We formalize the user node embedding update process via the graph convolution in the $l$-th layer of GNN for illustration as follows:
\begin{align}
    {\bf h}^l_u  = f_{conv}({\bf h}^{l-1}_p,{\bf h}^{l-1}_u; {\bm \theta}^l_{conv}), \forall p \in {\mathcal{N}(u)} ,\label{eq:gnn-conv}
\end{align}
where $f_{conv}$ represents the graph convolution operation and $\forall p \in {\mathcal{N}(u)}$ denotes any item $p$ belonging to the one-hop neighborhood of user $u$. Note that $\textbf{h}^0_u = \textbf{h}_u$ and $\textbf{h}^0_i = \textbf{h}_i$.
The refined user embedding $\textbf{h}^L_u$ and the item embedding $\textbf{h}^L_i$ outputted by the final layer of the GNN model are used to represent each user and item in a bipartite graph, and then typically we use the inner product over the user embedding ${\bf h}^L_u$ and item embedding ${\bf h}^L_i$ to make the recommendation prediction.

Overall, these GNN-based recommendation methods aim at learning the user/item embeddings $\bf H$ and the GNN model parameters ${\bm \Theta}_{GNN}$ simultaneously. Since interaction graphs from different recommendation tasks have distinct sets of users and items, apparently it is not feasible to transfer the learned embeddings, i.e., $\textbf{H}$.  

\begin{figure}[t]
    \centering
    \includegraphics[width=0.8\linewidth]{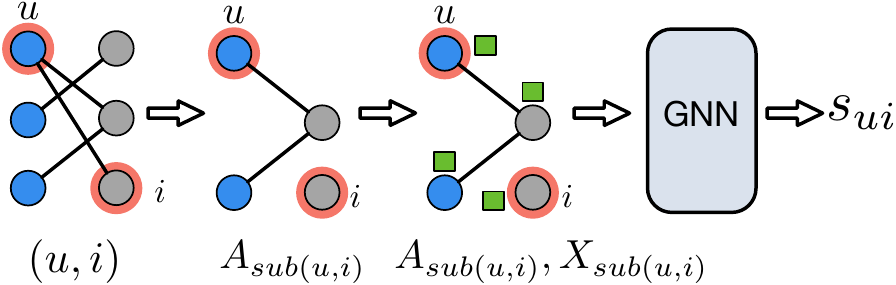}
    \caption{The proposed GNN recommendation method based on localized collaborative filtering. Given a target user $u$ and a target item $i$ , we first extract a local graph at these two target nodes from the user-item interaction graph, of which the structure is denoted as ${\bf A}_{sub(u,i)}$. Next, we generate a positional attribute for each node based on their minimum distances towards the target nodes, and these node attributes are denoted as ${\bf X}_{sub(u,i)}$ . Then, we can leverage a GNN model to get the graph representation for this attributed graph. Finally, we can compute a recommendation score $s_{ui}$ based on the local graph representation via a scoring function to make the final prediction for the given user and item pair.  }
    \label{fig:lgcf}
\end{figure}

In this paper, in order to ease pre-training for GNN-based recommendations, we propose a new GNN recommendation method based on localized collaborative filtering (or meta-LGNN). The intuition is that the key collaborative filtering information for recommendation is encoded in the historical interactions. Given a specific user and item pair, their key collaborative information is also encoded in their historical interactions. Thus it is reasonable to make a recommendation prediction based on the local structure consisting of their historical interactions. The framework of meta-LGNN is shown in Figure~\ref{fig:lgcf}. For a target user and item pair, we first extract a local graph at these two target nodes from the user-item interaction graph. Next, we generate a positional attribute for each node based on their minimum distances towards the target nodes. Then, we can leverage a GNN model to get the graph representation for this attributed graph. Finally, we can compute a recommendation score based on the local graph representation via a scoring function to make the final prediction for the given user and item pair.  

Given a target user $u$ and a target item $i$ in a user-item interaction graph $G$, we perform two random walks~\cite{pearson1905problem} starting from these two nodes on $G$, separately. Note that the random walk strategy we adopt in our work is with the restarting mechanism, and thus this process can be regarded as a neighbor sampling process, where we can get two sampled neighboring node sets $\mathcal{N}_u$ and $\mathcal{N}_i$ for $u$ and $i$, respectively. Next, we merge these two node sets to get an overall node set $N_{(u,i)}=\mathcal{N}_u \cup \mathcal{N}_i$, and finally we can get a local graph $G_{sub(u,i)}$ based on the set $N_{(u,i)}$ to reveal the local structure around the given pair $(u, i)$. Note that the reason why we only sample some neighboring nodes is to ensure the scalability of our model on large graphs. Apart from graph structure ${\bf A}_{sub(u,i)}$, we also need its node attributes ${\bf X}_{sub(u,i)}$, so that we can use a GNN model $ \mathcal{G}(\cdot;{\bm \Theta})$ to get a meaningful representation for $G_{sub(u,i)}$. For the model transferability, we proposed to use Double-Radius Node Labeling (DRNL)~\cite{zhang2018link} to generate a set of positional attributes for each user or item in a graph, rather than to use 
the existent node attributes, which are very likely to miss or be different across various recommendation tasks. DRNL labels each node based on its minimum distances towards the target user $u$ and the target item $i$ on the local graph $G_{sub(u,i)}$. Given a user/item node $t$, its positional attribute $x_t$ can be calculated as:
\begin{align}
    x_t = 1+\min(d_u,d_i)+(d/2)^2
\end{align}
\noindent where $d_u$ denotes the minimum distance between $u$ and $t$, and $d_i$ denotes the minimum distance between $i$ and $t$. $d=(d_u+d_i)$ is the sum of two minimum distances. Note that we denote $x_u =1$ and $x_i = 1$, so that we can distinguish the target user and item from other nodes. Now we have the graph structure and the node positional attributes of $G_{sub(u,i)}$, thus, we can use a GNN model $ \mathcal{G}(\cdot;{\bm \Theta})$ to encode it into a meaningful representation. The node embedding update process in each GNN layer also follows Eq.~\ref{eq:gnn-conv}. we use ${\bf H}^L_{sub(u,i)}\in \mathbb{R}^{n \times d}$ to denote the user-item embedding matrix outputted by the final GNN layer, where $L$ is the number of GNN layers, $n$ is the number of users and items in $G_{sub(u,i)}$ and $d$ denotes the embedding dimension. Then, the GNN model leverages a pooling operation to get the graph representation for $G_{sub(u,i)}$ as follows: 
\begin{align}
    {\bf h}_{(u,i)} = pool({\bf A}_{sub(u,i)}, {\bf H}^L_{sub(u,i)}).\label{eq:gnn-pool}
\end{align}
Note that the positional attributes are used as the input node features, i.e., ${\bf H}^0_{sub(u,i)} = {\bf X}_{sub(u,i)}$, which means that there is no need to learn the user/item embeddings. Finally, a score function is applied to compute a score for the user-item pair $(u,i)$ to make a prediction. Overall, meta-LGNN aims at only learning the GNN model parameters ${\bm \Theta}_{GNN}$ while eliminating the user/item embeddings $\bf H$.

\subsection{The ADAPT Pre-training}
In this subsection, we will introduce the overall pre-training process of the proposed ADAPT. In this pre-training phase, one key challenge is how to learn the common transferable knowledge across multiple interaction graphs and capture their differences simultaneously. To solve this challenge, we equip ADAPT with a GNN adaptor.  
Specifically, the GNN adaptor is able to adapt the meta-LGNN model to a customized GNN model particularly for the interaction graph in one recommendation task. Both the GNN adaptor and the meta-LGNN model are optimized in the pre-training process. Next, we detail the GNN adaptor component, the adaption process and the pre-training process.

\begin{figure*}[t]
    \centering
    \includegraphics[width=1.0\linewidth]{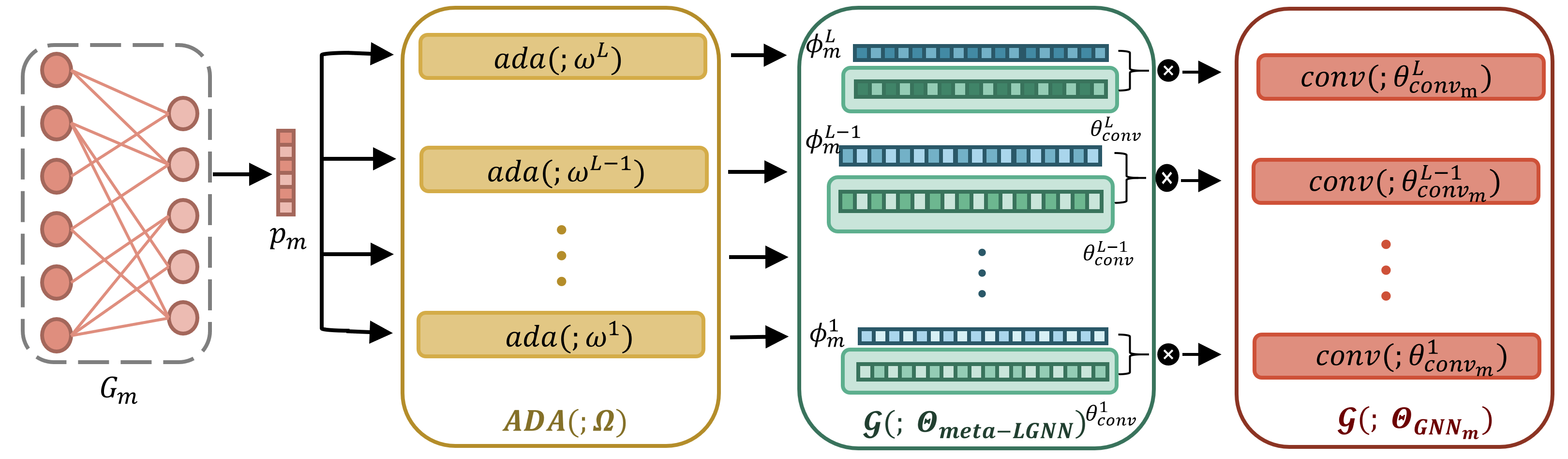}
    \caption{The adaptation process.  Given a specific graph $G_m$, the GNN adaptor takes its graph property vector ${\bf p}_m$ as input, and outputs customized adapting parameters ${\bf \Phi}_m$ for $G_m$. Suppose that the meta-LGNN model consists of $L$ GNN layers, we can denote its parameter as $ {\bm \Theta}_{Meta-LGNN} = \{{\bm \theta}^1_{conv},\cdots,{\bm \theta}^L_{conv}\}$, where ${\bm \theta}^l_{conv} \in \mathbb{R}^{d^{l-1}\times d^l}$ and $d^l$ denotes the dimension of node embeddings outputted by $l$-th GNN layer. The GNN adaptor will generate $L$ adapting parameters corresponding to these GNN layers for $G_m$, which can be denoted as $ {\bm \Phi}_m = \{{\bm \phi}^1_m,\cdots,{\bm \phi}^L_m\}$, where ${\bm \phi}^l_m\in \mathbb{R}^{2d^{l-1}d^l}$. For the graph convolution in the $l$-th GNN layer of the customized GNN model for graph $G_m$, its parameters ${\bm \phi}^l_m$ are adapted via ${\bm \phi}^l_m$ on ${\bm \theta}^l_{conv}$. }
    \label{fig:adaptation}
\end{figure*}
\subsubsection{The GNN Adaptor}\label{adaptormodel}
The goal of the GNN adaptor is to generate customized adapting parameters to adapt the meta-LGNN model for a given graph. Given a specific graph $G_m$, the GNN adaptor takes its graph property vector ${\bf p}_m$ as input, and outputs customized adapting parameters ${\bf \Phi}_m$ for $G_m$. In this work, we utilize some normalized graph structural properties such as graph size, graph density, and degree assortativity coefficient. As discussed before, suppose that the meta-LGNN model consists of $L$ GNN layers, we can denote its parameter as $ {\bm \Theta}_{Meta-LGNN} = \{{\bm \theta}^1_{conv},\cdots,{\bm \theta}^L_{conv}\}$, where ${\bm \theta}^l_{conv} \in \mathbb{R}^{d^{l-1}\times d^l}$ and $d^l$ denotes the dimension of node embeddings outputted by $l$-th GNN layer. The GNN adaptor will generate $L$ adapting parameters corresponding to these GNN layers for $G_m$, which can be denoted as $ {\bm \Phi}_m = \{{\bm \phi}^1_m,\cdots,{\bm \phi}^L_m\}$, where ${\bm \phi}^l_m\in \mathbb{R}^{2d^{l-1}d^l}$. The GNN adaptor can be modelled using any functions of ${\bf p}_m$. Specifically, for the graph convolution in the $l$-th GNN layer for $G_m$ , the GNN adaptor generates its corresponding adapting parameters as follows: 
\begin{align}
    {\bm \phi}^l_m = ada({\bf p}_m;\bm{\omega}^l), \label{eq:adaptor_l}
\end{align}

where $\bm{\omega}^l$ denotes the parameters of the GNN adaptor function $ada()$ corresponding to the graph convolution in the $l$-th GNN layer. In our work, we implement $ada()$ as a feed-forward neural network. Overall, we can summarize the adapting parameters generation process for $G_m$ as follows:
\begin{align}
    {\bm \Phi}_m = ADA({\bf p}_m;\bm{\Omega}), \label{eq:adaptor_l}
\end{align}
where $ADA()$ consists of all adaptor functions for different GNN layers and  $\bm{\Omega} = \{\bm{\omega}^1,\cdots,\bm{\omega}^L\}$ represents the parameters for all the GNN adaptors.
\subsubsection{The Adaptation Process}\label{metagnnmodel}

In our work, the meta-LGNN Model can be arbitrary GNN models including GCN~\cite{kipf2016semi}, GIN~\cite{xu2018powerful} and etc~\cite{ying2018hierarchical,ma2019graph,gao2019graph}. The meta-LGNN Model can be directly applied on a given the user-item interaction graph.

However, as aforementioned, there exist both similarities and differences among the user-item interaction graphs in different recommendation tasks. Thus to preserve similarities and differences simultaneously, we do not directly apply the same meta-LGNN model to all the recommendation graphs in the pre-training process. Instead, we utilize the GNN adaptor to generate a customized GNN model for each recommendation graph based on the meta-LGNN model and its graph properties. The adaptation process is illustrated in Figure~\ref{fig:adaptation}. Specifically, for the graph convolution in the $l$-th GNN layer of the customized GNN model for graph $G_m$, its parameters are adapted as follows:
\begin{align}
{\bm \theta}^l_{conv_m} ={\bm \theta}^l_{conv} \diamond {\bm \phi}^l_m,    
\end{align}
where $\diamond$ denotes an adaptation operation. In the current proposed model, we adopt FiLM~\cite{perez2018film} as the adaptation operation. Specifically, we split the adapting parameters ${\bm \phi}^l_m\in \mathbb{R}^{2d^{l-1}d^l}$ into two parts and reshape them as ${\bm \gamma}^l_m\in \mathbb{R}^{d^{l-1} \times d^l}$ and ${\bm \beta}^l_m\in \mathbb{R}^{d^{l-1} \times d^l}$. Then, we can formalize the adapting process of ${\bm \theta}^l_{conv}$ for $G_m$ as follows:
\begin{align}
{\bm \theta}^l_{conv_m} ={\bm \theta}^l_{conv} \odot {\bm \gamma}^l_m + {\bm \beta}^l_m, \label{eq:film}   
\end{align}
where $\odot$ denotes the element-wise multiplication between two matrices. By applying the adaptation operation on $L$ GNN layers of the meta-LGNN model, we can get the customized GNN model $ \mathcal{G}(\cdot;{\bm \Theta}_{GNN_m})$ for $G_m$, where ${\bm \Theta}_{GNN_m} = \{{\bm \theta}^1_{conv_m},\cdots,{\bm \theta}^L_{conv_m}\}$. The overall adaptation for $G_i$ can be summarised as:

\begin{align}
{\bm \Theta}_{GNN_m} ={\bm \Theta}_{Meta-LGNN} \Diamond {\bm \Phi}_m.
\end{align}

Then, we utilize the customized GNN model to generate graph embedding for the local graph $G_{sub(u,i)}$ extracted from $G_m$ for any user-item pair $(u,i)$ as follows:

\begin{align}
    {\bf h}_{u,i} = pool({\bf A}_{sub_{u,i}},\mathcal{G}({\bf A}_{sub_{u,i}}, {\bf X}_{sub_{u,i}}; {\bm \Theta}_{GNN_m})), \nonumber \\\forall (u,i) \in {E_m} \label{eq:ada-GNN}
\end{align}

\subsubsection{The Pre-training Process}~\label{pretrainloss}
In the pre-training phase, suppose that there are $N$ user-item interaction graphs $\{G_1,G_2,\cdots,G_N\}$ from different recommendation tasks. For any graph $G_m$, we can generate a customized GNN model $\mathcal{G}(; {\bm \Theta}_{GNN_m})$ based on the meta-LGNN model and graph properties ${\bf p}_m$ via the GNN adaptor. For any user-item pair $(u,i)\in {E_m}$, we utilize $\mathcal{G}(; {\bm \Theta}_{GNN_m})$ to compute a graph representation ${\bf h}_{u,i}$ for its corresponding local graph  $G_{sub_{u,i}}$, and then we use a scoring function to compute a recommendation score for $(u,i)$ based on ${\bf h}_{u,i}$ :
\begin{align}
     s_{(u,i)} = score({\bf h}_{(u,i)};\bm \delta)=\sigma({\bf h}_{(u,i)} \cdot {\bm \delta}),\label{eq:score}
\end{align}
where $\bm \delta$ is the linear transformation weights to be learned in the scoring function. We adopt the pairwise BPR loss~\cite{rendle2012bpr} in ADAPT. The BPR loss is one of the most popular objective functions in recommendation tasks, which measures the relative order of the positive node pairs and negative node pairs. The positive node pairs are user-item interactions observed in graphs, while the negative node pairs are non-existent user-item interactions generated by negative sampling. Specifically, BPR assumes that the recommendation score of the positive node pairs should be higher than the corresponding negative ones. The objective function $\mathcal{L}_{pre}$ of our model in the pre-training phase is formulated as follows:
\begin{align}
   \min_{\bm{\Omega},{\bm \Theta}_{Meta-LGNN},{\bm \delta}}\sum_{(u,i,i^{'})\in \mathcal{O}} -\ln\sigma( s_{(u,i)}- s_{(u,i^{'})}),\label{eq:pretrain-loss}
\end{align}
where $\mathcal{O}=\bigcup\limits_{m} \{(u,i,i^{'})|(u,i)\in E_m, (u,i^{'})\in E^-_m\}$ denotes the pre-training data for the graph set  $\{G_1,G_2,\cdots,G_N\}$. $E_m$ represents the existent interaction set between users and items in $G_m$ and $E^-_m$ is the non-existent interaction set. The overall pre-training process is summarized in Algorithm~\ref{alg:pre-train}. Given a set of interaction graphs used for pre-training $\{G_1,G_2,\cdots,G_N\}$ , batch size $b$ and the amount of interactions sampled for pre-training $N_{samples}$, we first randomly initialize the meta-LGNN model, the GNN adaptor and the scoring function. Next, in each training epoch (from line 3 to line 12), a random graph $G_m$ is sampled from $\{G_1,G_2,\cdots,G_N\}$ and the GNN adaptor generates its corresponding adaptation parameters ${\bm \Phi}_m$ based on its structure properties ${\bf p}_m$, and then the customized GNN model $\mathcal{G}(; {\bm \Theta}_{GNN_m})$ is generated especially for $G_m$. Furthermore, we randomly sample $b$ user-item interactions from the pre-training graph $G_m$ and generate their negative counterparts, and then the meta-LGNN model, the GNN adaptor and the scoring function are updated via minimizing the BPR loss on these samples. The training process will continue until the objective function is converged.

\begin{algorithm}[t]
	\caption{The Pre-training Process of ADAPT}
	\label{alg:pre-train}
	\begin{algorithmic}[1]
	\REQUIRE ~~\\
	A set of interaction graphs from different recommendation tasks: $\{G_1,G_2,\cdots,G_N\}$;\\
	Batch size $b$; \\
	Sample amount $N_{samples}$ (Note that $N_{samples} < \sum_m |E_m| $.)
	\ENSURE ~~\\
	The meta-LGNN model $\mathcal{G}(;{\bm \Theta}_{Meta-LGNN})$;\\
	the GNN adaptor $ADA(;\bm{\Omega})$;\\
	The scoring function $score(;\bm \delta)$
	\BlankLine
	\STATE Initialize ${\bm \Theta}_{Meta-LGNN}$, $\bm{\Omega}$ and $\bm \delta$ randomly;
	
	\WHILE{\textnormal{not converged}}
	    \FOR{batch=1,..,$\frac{N_{samples}}{b}$}
		    \STATE Sample a graph $G_m$ from $\{G_1,G_2,\cdots,G_N\}$;
		    
		    \STATE Compute ${\bm \Phi}_m = ADA({\bf p}_m;\bm{\Omega})$;
		    
		    \STATE Compute ${\bm \Theta}_{GNN_m} ={\bm \Theta}_{Meta-LGNN} \Diamond {\bm \Phi}_m$; 
		    
		    \STATE Sample $b$ user-item interactions $(u,i)\in E_m$;
		    
            \STATE Sample $b$ negative interactions $(u,i^{'})\notin E_m$;
            
            \STATE Compute $loss = \sum -\ln\sigma(score_{(u,i)}- score_{(u,i^{'})}) $
            
            \STATE Compute $grad = backward (loss)$
            
            \STATE Update( ${\bm \Theta}_{Meta-LGNN},\bm{\Omega},\bm \delta$; $grad$)
        \ENDFOR
		
    \ENDWHILE
	\end{algorithmic}
\end{algorithm}

\subsection{The ADAPT Fine-tuning}
\begin{figure}[t]
    \centering
    \includegraphics[width=0.7\linewidth]{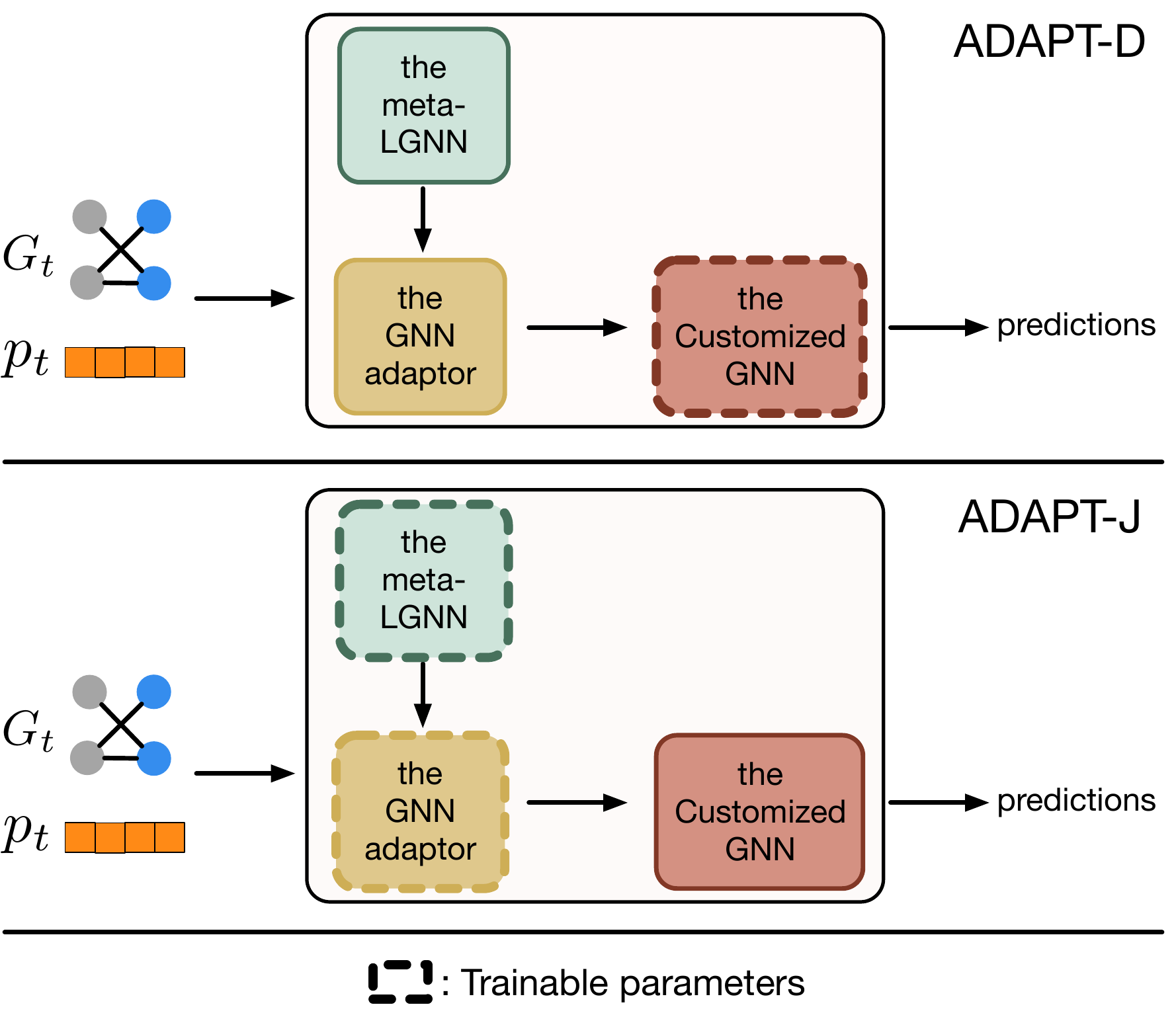}

    \caption{The fine-tuning strategies. We design two fine-tuning strategies -- \textit{direct fine-tuning} and \textit{joint fine-tuning}, which are denoted as \textit{ADAPT-D} and \textit{ADAPT-J}, separately. In  \textit{ADAPT-D}, we generate the customized GNN model ${\bm \Theta}_{GNN_t}$ based on the input graph via the GNN adaptor and the meta-LGNN, and we fine-tune the customized GNN model ${\bm \Theta}_{GNN_t}$; In \textit{ADAPT-J}, we fine-tune both the meta-LGNN model ${\bm \Theta}_{Meta-LGNN}$ and the GNN adaptor $\bm \Omega$.}
    \label{fig:finetune}
\end{figure}
\subsubsection{The Fine-tuning Objective Function }
In the fine-tuning phase, we also adopt the BPR loss. However, in this phase, we aim at optimizing the parameters of the customized GNN model. The formulation is listed below:
\begin{align}
    \min_{{\bm \Theta}_{GNN_t}}\sum_{(u,i,i^{'})\in {E_t^{'}}} -\ln\sigma(\hat y_{(u,i)}-\hat y_{(u,i^{'})}),\label{eq:finetune-loss}
\end{align}
where $E_t^{'} = \{(u,i,i^{'})|(u,i)\in E_t, (u,i^{'})\in E^-_t\}$ denotes the training data for the target graph $G_t$. $E_t$ represents the set of existent interactions between users and items in $G_t$ and $E^-_t$ is the non-existent interaction set generated manually. 

\subsubsection{The Fine-tuning Strategies}
With the GNN adaptor and the meta-LGNN model from the pre-training phase, we can conduct model fine-tuning specially for a target recommendation graph $G_t$. We design two fine-tuning strategies and illustrate them in Figure~\ref{fig:finetune}. 
\begin{itemize}
\item The first fine-tuning strategy is named as \textit{direct fine-tuning}. As shown in the top subfigure of Figure~\ref{fig:finetune}, given a target downstream graph $G_t$, we first compute the customized adapting parameters $\bm{\Phi}_t = ADA(p_t;\bm{\Omega})$ based on its property vector $p_t$ via the GNN adaptor $ADA(;\bm{\Omega})$, then generate the customized GNN model ${\bm \Theta}_{GNN_t} ={\bm \Theta}_{Meta-LGNN} \Diamond {\bm \Phi}_t$. Finally, we fine-tune ${\bm \Theta}_{GNN_t} $ by optimizing the objective function described in Eq.~\ref{eq:finetune-loss}. We denote ADAPT with this fine-tuning strategy as \textit{ADAPT-D}. 
\item The second fine-tuning strategy is named as \textit{joint fine-tuning}. As shown in the bottom subfigure of Figure~\ref{fig:finetune}, given a target downstream graph $G_t$, we fine-tune both the meta-LGNN model ${\bm \Theta}_{Meta-LGNN}$ and the GNN adaptor $\bm \Omega$ on $G_t$. This fine-tuning strategy is inspired by test-time training~\cite{sun2020test}, which is proposed to fine-tune the model for a test sample in the test process so that the model can be better adapted for the test data. We use \textit{ADAPT-J} to indicate ADAPT with the joint fine-tuning strategy. 
\end{itemize}

\section{Experiments}~\label{sec:experiment}
In this section, we conduct extensive experiments to validate the effectiveness of the proposed ADAPT. We first introduce the experimental settings. Then we illustrate how the sparsity of the recommendation graphs affects the performance of existent GNN-based recommendation methods. Next, we evaluate the performance of ADAPT and representative baselines on various real-world datasets. We conduct the ablation study to understand the importance of the GNN adaptor. Finally, we investigate the impact of graph sparsity and the number of pre-training graphs on the performance of ADAPT. 

\subsection{Experimental Settings}
\begin{figure}[t]
  \centering
  \subfloat[60 pre-training graphs from Tianchi \label{fig:pre-61}]{{\includegraphics[width=0.5\linewidth]{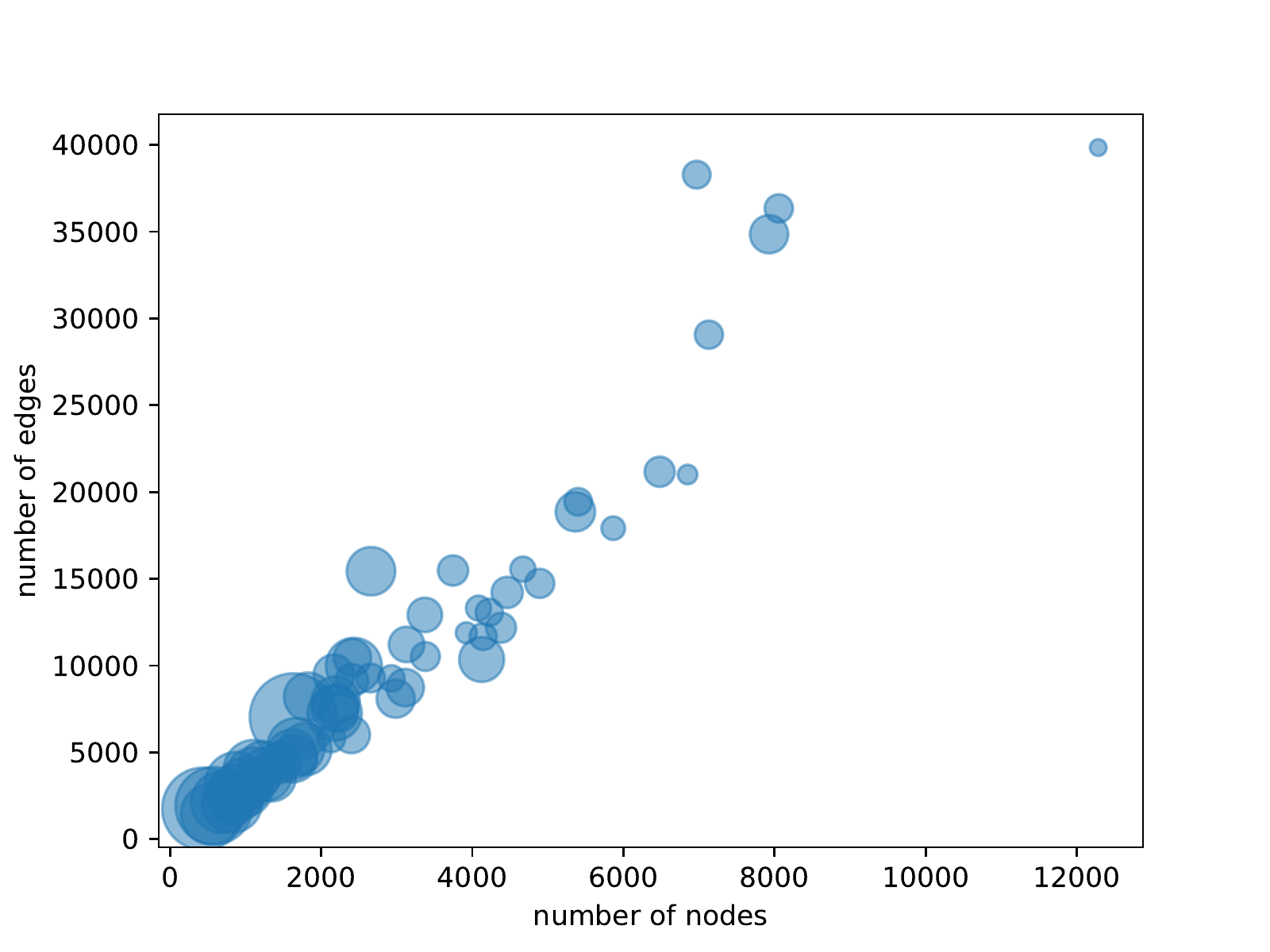} }}
    \subfloat[6 pre-training graphs from MovieLens  \label{fig:pre-war}]{{\includegraphics[width=0.5\linewidth]{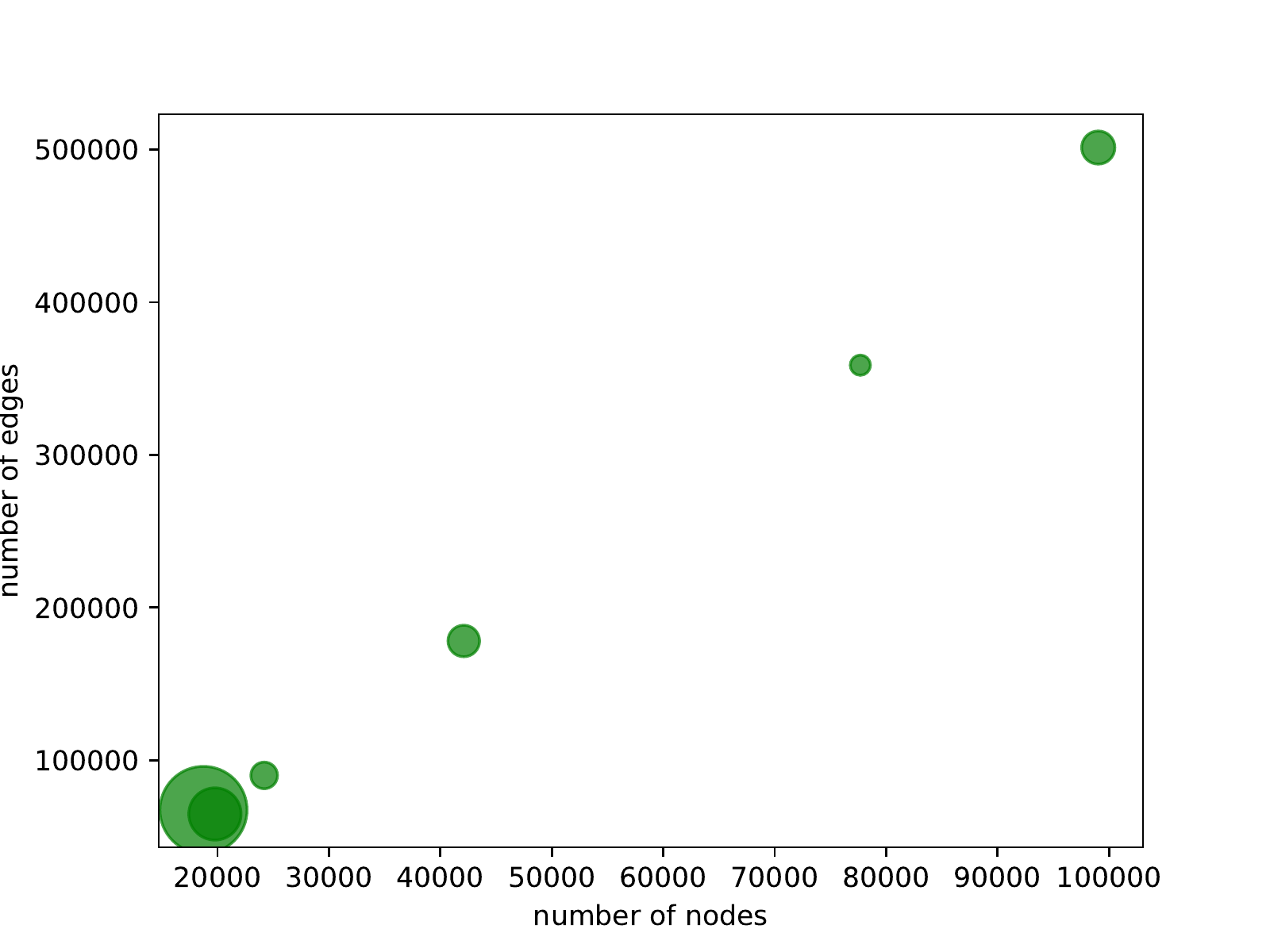} }}
\caption{Properties of pre-training graphs. Note that each circle represents a graph and its size denotes the density of the graph.}\label{fig:pre-pro}
\end{figure}

In this subsection, we introduce the experimental settings including datasets, baselines and evaluation metrics. 
\subsubsection{Datasets}

In this work, we conduct experiments on datasets from two real-world applications: Tianchi~\cite{tianchi} and MovieLens~\cite{movielens}. Specifically, we construct numerous user-item interaction graphs from different recommendation scenarios based on the item category from these two applications. For each application, we select some interaction graphs as pre-training graphs and some of them as the target downstream graphs. The statistics of the pre-training graphs are demonstrated in Figure~\ref{fig:pre-pro} and these of the downstream graphs are summarized in Table~\ref{table:datasets}. As shown in Figure~\ref{fig:pre-pro}, we select 60 interaction graphs from Tianchi and 6 interaction graphs from MovieLens for pre-training, and these graphs show very diverse properties. We briefly introduce the Tianchi dataset and the MovieLens dataset as follows:

\begin{table}[]
\caption{Statistics of the downstream graphs from Tianchi and MovieLens.}

\centering
 \setlength{\tabcolsep}{6mm}{
\begin{tabular}{c|c|c|c}
\hline
Dataset & \#Users & \#Items & \#Edges\\ \hline

\textbf{Tianchi-174490}& 	2,267& 	285	& 3,826	\\ \hline
\textbf{Tianchi-61626}	& 2,305	& 764& 	3,270\\ \hline
\textbf{Tianchi-3937919}	& 1,846& 	158& 	2,980\\ \hline
\textbf{Tianchi-2798696}  & 2,579    &94      & 2,988\\ \hline
\textbf{MovieLens-War}	& 2,970	& 89	& 4,240\\ \hline

\end{tabular}}
\label{table:datasets}
\end{table} 
\begin{itemize}
    \item  \textbf{Tianchi}: It is a user behavior dataset from \textit{Alibaba} (one of the biggest on-line shopping platforms in China), which consists of millions of user-item interactions, such as clicking, liking and purchasing. Each interaction record includes user ID, item ID, behavior type, item category ID and interaction timestamp. In our work, we use Tianchi-$ID$ to denote the user-item interaction graph whose items belong to the category $ID$. 
    \item \textbf{MovieLens}: This is a movie rating dataset from  \textit{MovieLens} (a popular movie recommendation website), which consists of user rating records for thousands of movies from different categories. Each rating record includes user ID, movie ID, rating, and movie category. In our work, we use MovieLens-$X$ to denote the user-movie interaction graph whose movies come from the category $X$.   
\end{itemize}

\subsubsection{Baselines}
In this subsection, we describe representative baseline methods from two groups. The first group includes GNN-based recommendation methods and a classic CF method. Particularly, we select NGCF and LightGCN since they are two of the most representative GNN-based recommendation methods, and LightGCN is one of the state-of-the-art models. MF is chosen because it is one of the most classic and popular recommendation methods. The second group includes existing pre-training methods for GNNs. Though there are a few pre-training methods for GNNs \cite{hu2019strategies,hu2020gpt,qiu2020gcc}, the majority of them are not designed specifically to recommendations. Thus, we adapt a recent method GCC for recommendations as the representative baseline since it only relies on the topological information and can be pre-trained on multiple graphs as the proposed ADAPT does. We have not chosen the pre-training strategies proposed in~\cite{hu2019strategies} because there are no node attributes or graph labels in our scenarios. Similarly, GPT-GNN~\cite{hu2020gpt} has not been included because it also requires node attributes and cannot be applied across multiple graphs. The details of these baselines are presented as follows: 
\begin{itemize}
    \item \textbf{NGCF}: NGCF~\cite{wang2019neural} proposes to encode high-order connectivity among user-item interaction graphs into user/item embeddings. Specifically, it designs propagation layers to aggregate information from connected nodes. The recommendation prediction is made based on the refined user and item embeddings.

    \item \textbf{LightGCN}: LightGCN~\cite{he2020lightgcn} is one of the state-of-the-art GNN-based recommendation methods. It tailors GNN particularly for recommendation tasks.
    Specifically, it eliminates feature transformation and nonlinear activation, and only includes neighborhood aggregation in GNNs for collaborative filtering.

    \item \textbf{MF}: MF~\cite{koren2009matrix} is one of the most classic and popular recommendation methods. It learns a set of latent user and item embeddings directly from user-item interactions, and then makes recommendation predictions based on the inner product of the user embedding and the item embedding.
    
    \item \textbf{GCC}: GCC~\cite{qiu2020gcc} is a self-supervised pre-training framework for GNNs. GCC designs a subgraph instance discrimination task as the pre-training task and it utilizes contrastive learning to empower GNNs to learn transferable structural representations across multiple graphs. Note that GCC is not specially designed for recommendation tasks. To apply GCC into the recommendation scenarios, we first pre-train a GCN model via the GCC framework on the pre-training graphs. Then we build a scoring function model specific to recommendations in the fine-tuning phase. The fine-tuning objective function is the same as ADAPT.
 
\end{itemize}
Note that since we do not focus on the cold-start problem, we do not include the pre-training work in~\cite{hao2021pre} as one baseline.

\subsubsection{Evaluation Metrics}
For the downstream recommendation tasks, we divide its corresponding user-item interactions into three sets: the training set, the validation set and the test set. Note that each of the validation set and the test set has $5$ percent of the total samples. To avoid the cold-start problem, we constrain that all the users and items in the validation set and the test set should exist in the training set. In the fine-tuning phase, we use the validation set to select the best model and then report its performance in the test set. For each user-item interaction in the validation or the test set, we generate 49 non-existent user-item interactions for the user. In the model evaluation, we first compute a recommendation score for each interaction, and then we rank these scores. We calculate the Hit Rate (HR) of the model prediction based on if the score of the real user-item interaction is on the top 5 among all the 50 user-item interactions. For each experiment, we average and report the model performance with $5$ seeds in terms of HR.

\subsubsection{Implementation Details}
In our work, we implement the meta-LGNN model as a 3-layer graph convolutional networks and the scoring function as a linear transformation layer. For the input of the GNN adaptor, we compute eight graph properties via the networkX package~\cite{hagberg2008exploring}, including the number of nodes, the number of edges, the user-item ratio, the graph density, the degree assortativity coefficient, robins-alexander clustering coefficient, the number of connected components and the global efficiency. Note that it is straightforward to include more graph properties. We use the Adam~\cite{kingma2014adam} optimizer. The batch size is set to be 256. The learning rate is chosen from $[0.0005, 0.001]$, and the dropout rate is chosen from $[0.0, 0.1, 0.2, 0.3]$.

\subsection{Preliminary Study}
\begin{figure}[t]
  \centering
  \subfloat[Tianchi-61626\label{fig:pre-61}]{{\includegraphics[width=0.4\linewidth]{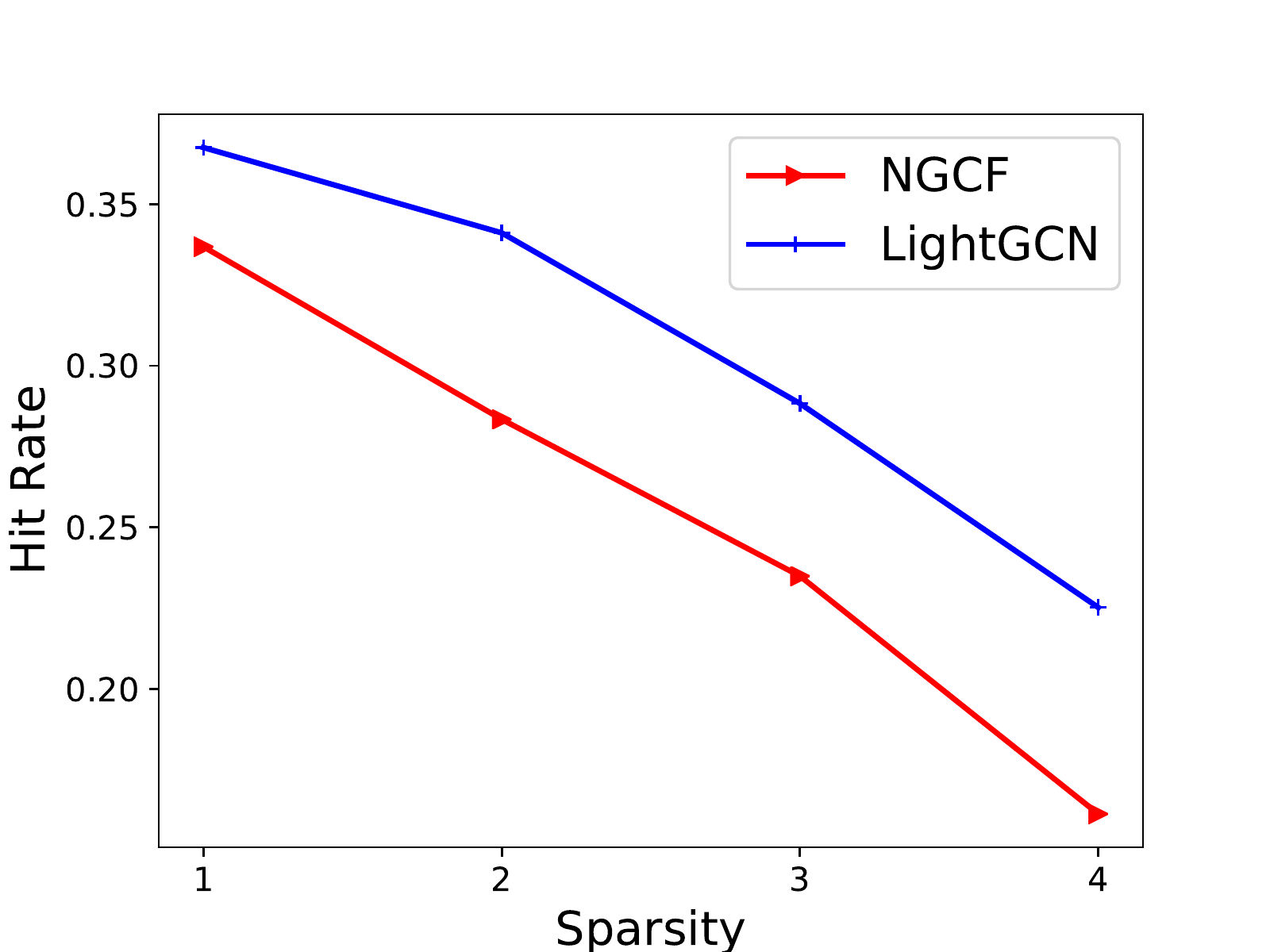} }}
    \subfloat[Tianchi-174490\label{fig:pre-war}]{{\includegraphics[width=0.4\linewidth]{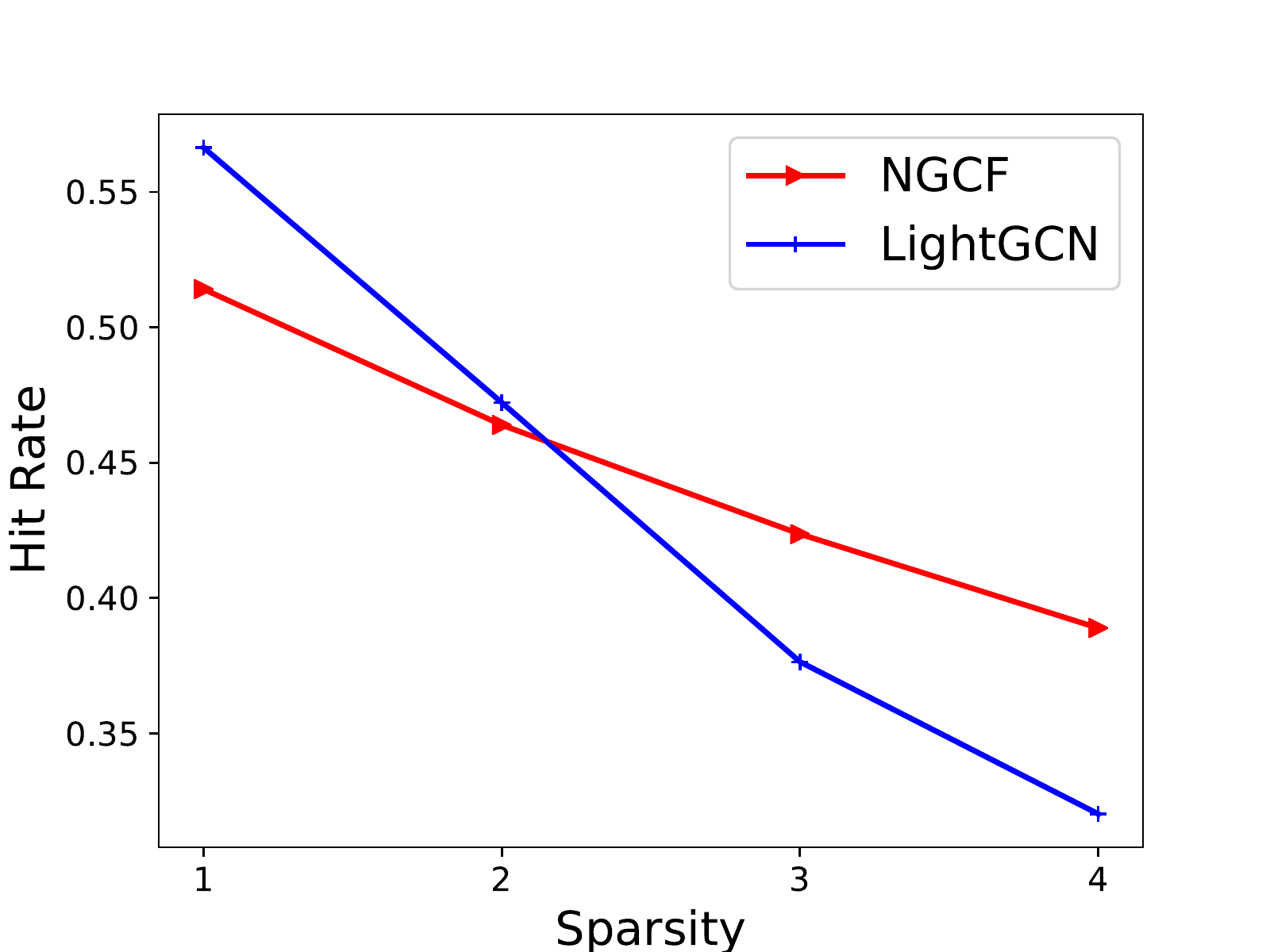} }}
\caption{The performance of NGCF and LightGCN vs. the graph sparsity on two Tianchi dataset datasets. (The reported performance is measured with HR in $\%$)}\label{fig:pre}
\end{figure}

In this subsection, we study how the performance of two representative GNN-based recommendation methods, i.e., NGCF and LightGCN, is affected by the sparsity of interaction graphs. Specifically, we increase the sparsity of interaction graphs by randomly removing parts of the training edges while fixing the test edges. The results are shown in Figure~\ref{fig:pre} where $x\in\{1,2,3,4\}$ (i.e., the x-axis) is used to denote the graph sparsity and a larger $x$ indicates a sparser graph. Both LightGCN and NGCF show a significant decreasing trend in terms of recommendation performance with the increase of graph sparsity. This empirically demonstrates that GNN-based recommendation methods suffer from data sparsity. This motivates us to take advantage of pre-training techniques to alleviate this problem. 

\begin{table*}[t]

\caption{The recommendation performance comparison with $60\%$ as training. (The reported performance is measured with HR in $\%$)}
\scalebox{0.8}{
\begin{tabular}{lcccc|c|ccc|c}
\toprule
  \multirow{2}{*}{\textbf{Datasets}}   & \multicolumn{4}{c}{\textbf{Training from Scratch}} & \multicolumn{4}{c}{\textbf{Pre-training \& Fine-tuning}} & \multirow{2}{*}{\makecell[c]{\textbf{Performance}\\ \textbf{ Gain}}}\\ 
  \cmidrule(r){2-5} \cmidrule(r){6-9} 
  & MF   & NGCF & LightGCN  & \textbf{Best} & GCC & ADAPT-D & ADAPT-J& \textbf{Best }\\\midrule
Tianchi-174490  & 33.51$\pm$3.71 & 46.39$\pm$3.07 & 47.22$\pm$3.08 & \textbf{47.22$\pm$3.08} & 37.17$\pm$15.6 & 60.94$\pm$1.46 & 58.64$\pm$2.85 & \textbf{ 60.94$\pm$1.46} & \textbf{+29.05\%} \\
Tianchi-61626 & 17.30$\pm$3.14 & 28.34$\pm$5.52 &34.11$\pm$3.52 &\textbf{34.11$\pm$3.52} & 33.37$\pm$3.4 & 50.06$\pm$3.18 & 49.81$\pm$3.16 &\textbf{50.06$\pm$3.18} & \textbf{+46.77\%}  \\
Tianchi-3937919 & 21.75$\pm$3.95 & 26.71$\pm$4.59 &40.2$\pm$3.31 &\textbf{40.2$\pm$3.31}& 35.57$\pm$12.12 &57.72$\pm$2.31 & 59.6$\pm$1.97& \textbf{59.6$\pm$1.97} & \textbf{+48.25\%} \\
Tianchi-2798696 &30.81$\pm$5.22 &35.97$\pm$2.71 &28.99$\pm$3.86 &\textbf{35.97$\pm$2.71} & 45.1$\pm$3.4 & 47.38$\pm$2.15 &46.84$\pm$0.28 &\textbf{47.38$\pm$2.15} & \textbf{+31.72\%} \\  \midrule
MovieLens-War & 62.94$\pm$4.25 & 63.41$\pm$3.06 &62.94$\pm$4.07&\textbf{63.41$\pm$3.06}& 64.93$\pm$5.13 & 73.93$\pm$1.18 &75.99$\pm$0.72&\textbf{75.99$\pm$0.72}& \textbf{+19.84\%}\\                                                                          
\bottomrule
\end{tabular}
}
\label{tab:main-performance}
\end{table*}

\begin{table*}[t]

\caption{The recommendation performance comparison with $40\%$ as training. (The reported performance is measured with HR in $\%$)}
\scalebox{0.8}{
\begin{tabular}{lcccc|c|ccc|c}
\toprule
  \multirow{2}{*}{\textbf{Datasets}}   & \multicolumn{4}{c}{\textbf{Training from Scratch}} & \multicolumn{4}{c}{\textbf{Pre-training \& Fine-tuning}} & \multirow{2}{*}{\makecell[c]{\textbf{Performance}\\ \textbf{ Gain}}}\\ 
  \cmidrule(r){2-5} \cmidrule(r){6-9} 
  & MF   & NGCF & LightGCN  & \textbf{Best} & GCC & ADAPT-D & ADAPT-J& \textbf{Best }\\\midrule
Tianchi-174490  & 30.73$\pm$4.35 & 42.36$\pm$2.96 & 37.64$\pm$3.24 & \textbf{42.36$\pm$2.96} & 33.40$\pm$1.36 & 57.70$\pm$2.49 & 60.07$\pm$3.2 & \textbf{ 60.07$\pm$3.20} & \textbf{+41.82\%} \\
Tianchi-61626 & 16.32$\pm$3.21 & 23.50$\pm$4.89 &28.83$\pm$4.48 &\textbf{28.83$\pm$4.48} & 32.27$\pm$2.32 & 49.94$\pm$3.75 & 52.39$\pm$1.27 &\textbf{52.39$\pm$1.27} & \textbf{+81.70\%}  \\
Tianchi-3937919 & 20.60$\pm$3.84 & 25.57$\pm$3.62 &35.44$\pm$1.70 &\textbf{35.44$\pm$1.70}& 34.9$\pm$17.0& 49.13$\pm$8.98 & 58.79$\pm$3.37& \textbf{58.79$\pm$3.37} & \textbf{+65.9\%} \\
Tianchi-2798696 &28.86$\pm$4.12 &33.29$\pm$1.98 &27.11$\pm$4.89 &\textbf{33.29$\pm$1.98} & 46.17$\pm$2.53 & 48.66$\pm$2.86 &46.58$\pm$2.10 &\textbf{48.66$\pm$2.86} & \textbf{+46.16\%} \\  \midrule
MovieLens-War & 61.14$\pm$4.52 &64.17$\pm$2.50& 56.78$\pm$4.86& \textbf{64.17$\pm$2.50}&  50.76$\pm$13.65 & 71.85$\pm$1.76 &71.47$\pm$1.18&\textbf{ 71.85$\pm$1.76}& \textbf{+11.97\%}\\                                                                          
\bottomrule
\end{tabular}
}
\label{tab:main-performance-sparser}
\end{table*}

\subsection{Performance Comparison}
We compare the proposed ADAPT and baselines described aforementioned on five datasets, including four item categories from Tianchi and one movie category from MovieLens. The recommendation methods are divided into two groups: 1) the \textit{training from scratch} group consists of MF, NGCF and LightGCN, which are directly trained from scratch on the target downstream graphs; and 2) the \textit{pre-training \& fine-tuning} group consists of GCC and the proposed ADAPT with two fine-tuning strategies, i.e., ADAPT-D and ADPAT-J. To ease the comparison, we also show the best performance of methods from these two groups, and compute the performance gain of the \textit{pre-training \& fine-tuning} group over the \textit{training from scratch} group. Note that in order to simulate the data scarcity scenarios in real-world recommendation applications, after the validation set and the test set are determined, we randomly remove some interactions from the remaining interactions under the constraint of introducing no isolated users or items. Specifically, we tailor two training sets: One retains 60 percent of the remaining interactions; and the other one consists of 40 percent interactions. The comparison results on these two scenarios are shown in Table~\ref{tab:main-performance} and Table~\ref{tab:main-performance-sparser}, respectively. We can make the following observations: 
\begin{itemize}
    \item NGCF and LightGCN often perform better than MF. This observation is consistent with previous observations in~\cite{wang2019neural,he2020lightgcn}. 
    \item  The proposed ADAPT frameworks can achieve great performance improvement over the recommendation models trained from scratch on the target downstream data. This demonstrates the effectiveness of ADAPT in alleviating the data scarcity problem in recommendation tasks.
    \item ADAPT achieves significantly better performance than GCC. Compared to GCC, ADAPT is designed specifically to the pretraining task for recommendations with the recommendation BPR objective and the adaptor to capture the differences among pre-training graphs. More investigations on the importance of the adaptor will be discussed in the following subsection. 
    \item Both the \textit{direct fine-tuning} strategy and the \textit{joint fine-tuning} strategy are effective. The \textit{joint fine-tuning} strategy is empirically shown to be more effective than the \textit{direct fine-tuning} strategy in most cases. This observation could indicate that fine-tuning the adaptor in the test time has the potential to benefit the performance of the downstream recommendation task. This observation is consistent with that in~\cite{sun2020test}.
    \item The performance gain of ADAPT from the best baseline performance is more significant under the $40\%$ than $60\%$. This observation suggests that pre-training is a promising solution to tackle the data sparsity problem in recommendations.
\end{itemize}

\subsection{Ablation Study}

\begin{table*}[t]

\caption{Ablation study of the GNN adaptor. (The reported performance is measured with HR in $\%$)}
 \setlength{\tabcolsep}{2mm}{
\begin{tabular}{lcccc}
\toprule
  \multirow{3}{*}{\textbf{Datasets}}& \multicolumn{3}{c}{\textbf{Pre-training \& Fine-tuning}}& \textbf{Training from Scratch} \\
   \cmidrule(r){2-5}
   & \multicolumn{2}{c}{60 pre-training graphs}& 1 pre-training graph &  \multirow{2}{*}{ADAPT-w/o-adaptor-scratch}\\
   
  & ADAPT-best & ADAPT-w/o-adaptor  & ADAPT-w/o-adaptor\\\midrule
Tianchi-174490 & 60.07$\pm$3.2 &  32.25$\pm$9.43 & 43.04$\pm$14.56 &53.92$\pm$10.58\\
Tianchi-61626 & 52.39$\pm$1.27 & 44.78$\pm$9.83& 43.81$\pm$6.36 & 49.57$\pm$3.98 \\ 
Tianchi-3937919 & 58.79$\pm$3.37 &38.12$\pm$7.29& 48.59$\pm$7.89 & 58.25$\pm$5.89\\
Tianchi-2798696 & 48.66$\pm$2.86 & 47.36$\pm$1.46& 45.64$\pm$1.50 & 47.25$\pm$2.15\\ \midrule
\textbf{Average}  &54.98 & 40.63 &45.27 & 52.24\\   \bottomrule
\end{tabular}
}

\label{tab:adaptor}
\end{table*}

We conduct an ablation study to investigate the effectiveness of the GNN adaptor. The results are shown in Table~\ref{tab:adaptor}. We use ADAPT-w/o-adaptor to denote the ADAPT framework without the GNN adaptor. We pre-train two model instances of ADAPT-w/o-adaptor with 60 pre-training graphs and 1 pre-training graph, respectively. We use ADAPT-w/o-adaptor-scratch to indicate that ADAPT-w/o-adaptor is trained from scratch with only the downstream graph. ADAPT-best denotes the ADAPT framework with any fine-tuning strategy that can achieve better performance. We can make the following observations:
\begin{itemize}
    \item The overall performance of ADAPT-best is significantly better than that of the ADAPT-w/o-adaptor and ADAPT-w/o-adaptor-scratch. This validates (1) the effectiveness of the GNN adaptor for pre-training and (2) the importance of pre-training.
    \item The ADAPT-w/o-adaptor instance pre-trained on 60 pre-training graphs performs even worse than the instance pre-trained on a single pre-training graph. This shows that it is necessary to capture the differences among multiple pre-training graphs, and the GNN adaptor in the proposed ADAPT can capturethese differences in the pre-training process.
    \item Both the ADAPT-w/o-adaptor instances cannot beat ADAPT-w/o-adaptor-scratch. This indicates that pre-training is not always helpful. It can even introduce performance degradation if not carefully designed. 
\end{itemize}

To further demonstrate the importance of pre-training, we check if ADAPT can transfer knowledge from the pre-training graphs to the downstream recommendation task. To achieve this goal, we directly use the models generated from the pre-training without fine-tuning. In particular, we compare the performance of three different variants of the proposed GNN recommendation method for localized collaborative filtering. They include a randomly-initialized model, the meta-LGNN model from the pre-training phase, and the customized-GNN model generated by the meta-LGNN model and the GNN adaptor especially for the target dataset. The results are demonstrated in Table~\ref{tab:diff-init}. Overall, the meta-LGNN model performs better than the randomly-initialized model, and the performance of the customized-GNN model is significantly better than that of the other two variants. These observations demonstrate that the meta-LGNN model learns some useful knowledge from the pre-training process, and the GNN adaptor effectively adapts the meta-LGNN model to the customized one. The customized-GNN model performs remarkably better than the randomly-initialized model. This provides direct evidence that ADAPT successfully transfers knowledge from pre-training graphs to the downstream recommendation task. 

\begin{table}[t]

\caption{Performance comparison of three different variants of the proposed GNN recommendation method for localized collaborative filtering. (The reported performance is measured with HR in $\%$)}

\begin{tabular}{lccc}
\toprule
  \textbf{Datasets}&random-ini& meta-LGNN & customized-GNN \\\midrule
Tianchi-174490 & 16.44 &  13.09& 28.17 \\
Tianchi-61626 &12.64 &16.20 & 26.99  \\ 
Tianchi-3937919 & 16.11 &12.89& 30.33 \\
Tianchi-2798696 & 11.68 & 19.33& 17.59\\ \midrule
\textbf{Average} &14.22 & 15.38 & 25.77 \\   \bottomrule
\end{tabular}
\label{tab:diff-init}
\end{table}

\subsection{Further Probing}
In this subsection, we further probe the proposed framework by exploring the following two problems: 1) how does the sparsity of the target downstream graph affect the ADAPT performance? and 2) how does the number of pre-training graphs influence the ADAPT performance? 

\subsubsection{Sparsity Analysis}

\begin{figure*}[ht]%
\vspace{0.1in}
    \centering

    \subfloat[Tianchi-174490 \label{fig:sp17}]{{\includegraphics[width=0.33\linewidth]{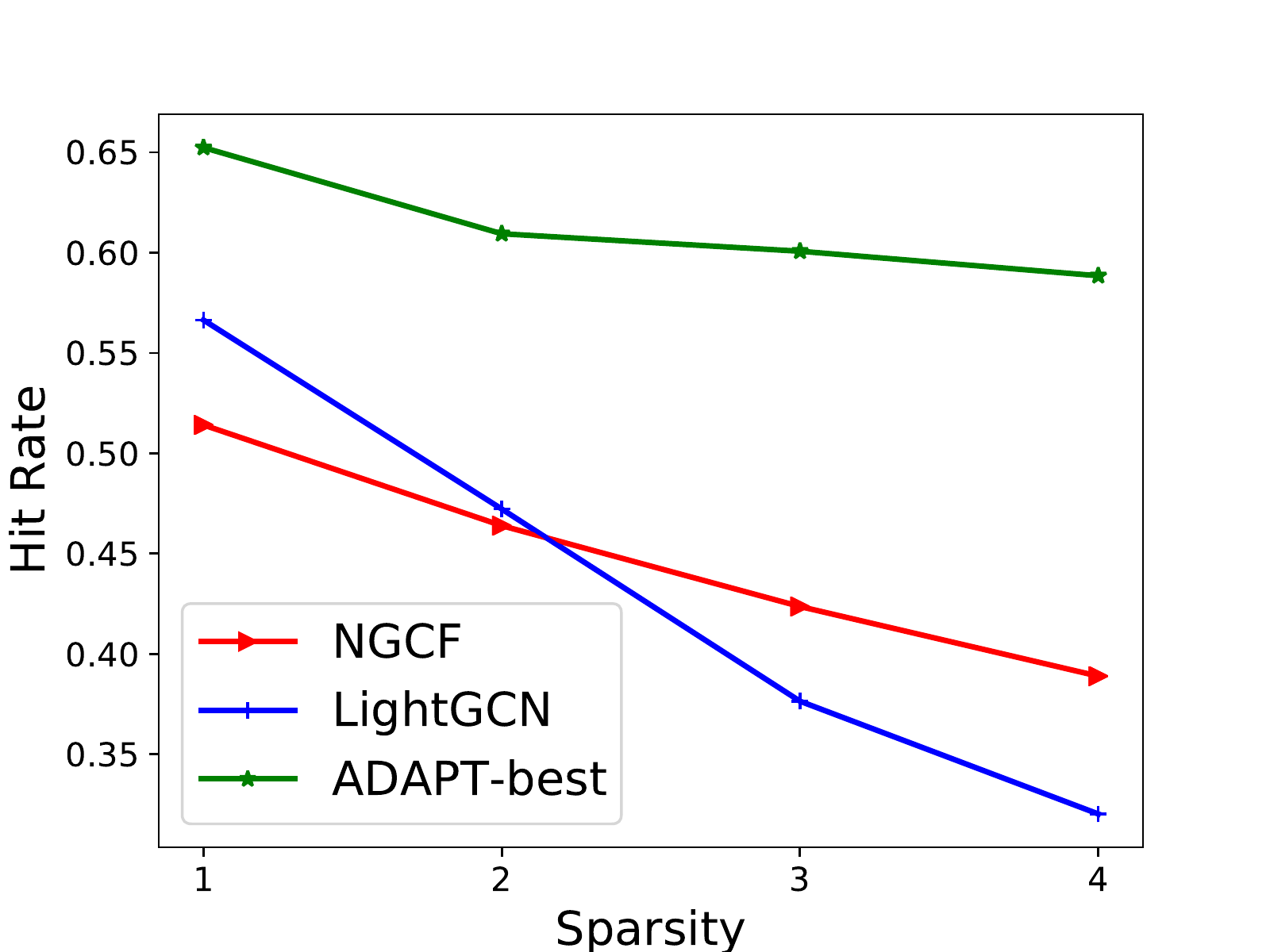}}}
    \subfloat[Tianchi-61626\label{fig:sp61}]{{\includegraphics[width=0.33\linewidth]{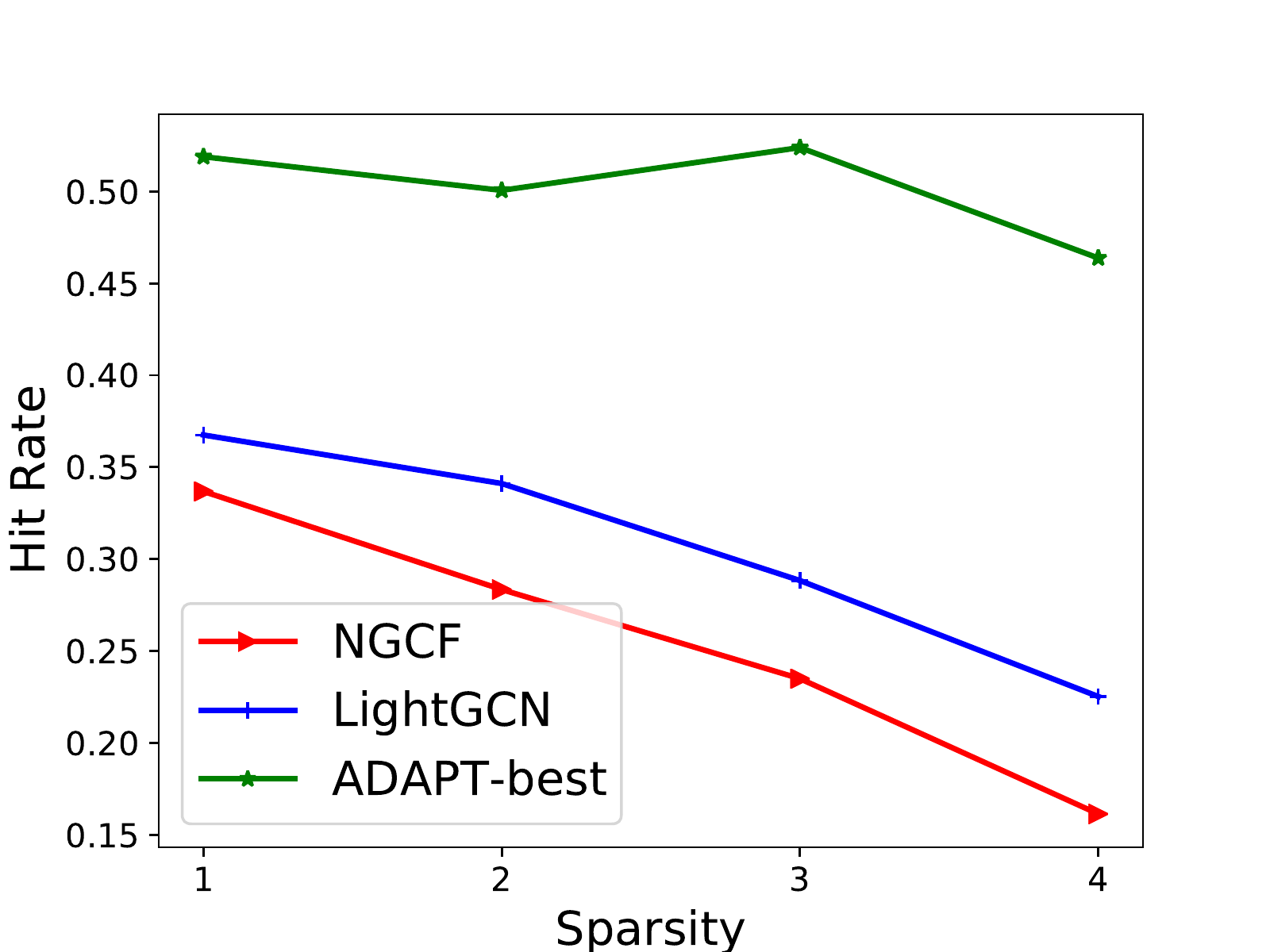} }}%
    \subfloat[MovieLens-War \label{fig:spwar}]{{\includegraphics[width=0.33\linewidth]{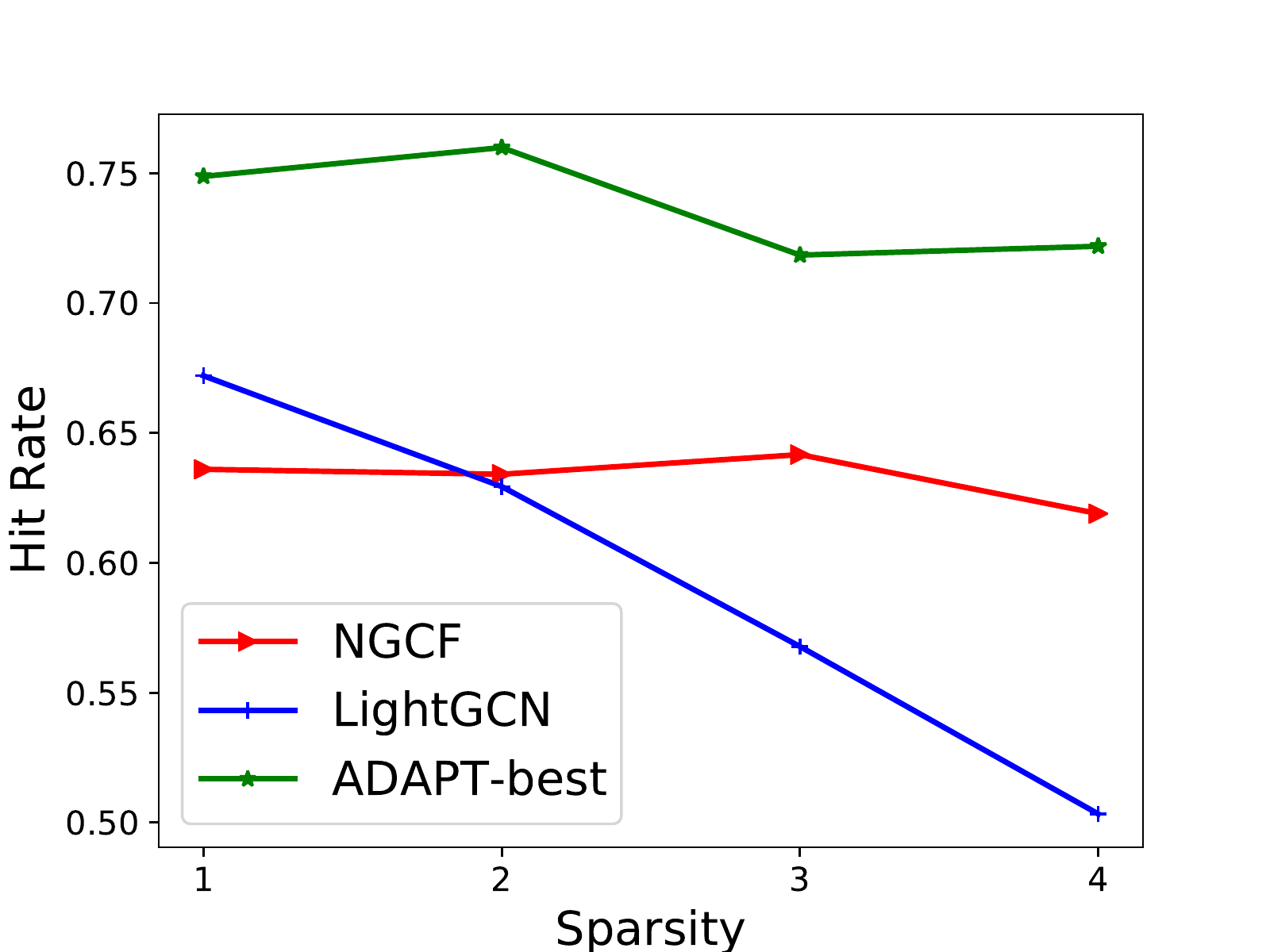}}}%

   \vspace{0.1in}
    \caption{The performance of the proposed ADAPT with varied sparsity of the target downstream graph. Note that the x-axis is used to denote the graph sparsity. The larger the sparsity is, the sparser the target downstream graph is.}%
    \label{fig:case-study}
\vspace{0.1in}
\end{figure*}
In Figure~\ref{fig:case-study}, we show how the performance of ADAPT and two GNN-based recommendation methods, NGCF and LightGCN, changes with the change of graph sparsity. Specifically, we gradually increase the graph sparsity by randomly removing some interactions from the training set under the constraint of introducing no isolated users or items, and meanwhile, the test set and validation set remain unchanged. It is observed that the proposed ADAPT performs much more stable with the increase of the graph sparsity, compared to NGCF and LightGCN. This observation further demonstrates the appealing of pre-training to mitigate the data sparsity problem in recommendations. 

\subsubsection{The Number of Graphs for Pre-training}
\begin{figure}[ht]%

 \centering

 \subfloat[Tianchi-174490 \label{fig:17-num}]{{\includegraphics[width=0.4\linewidth]{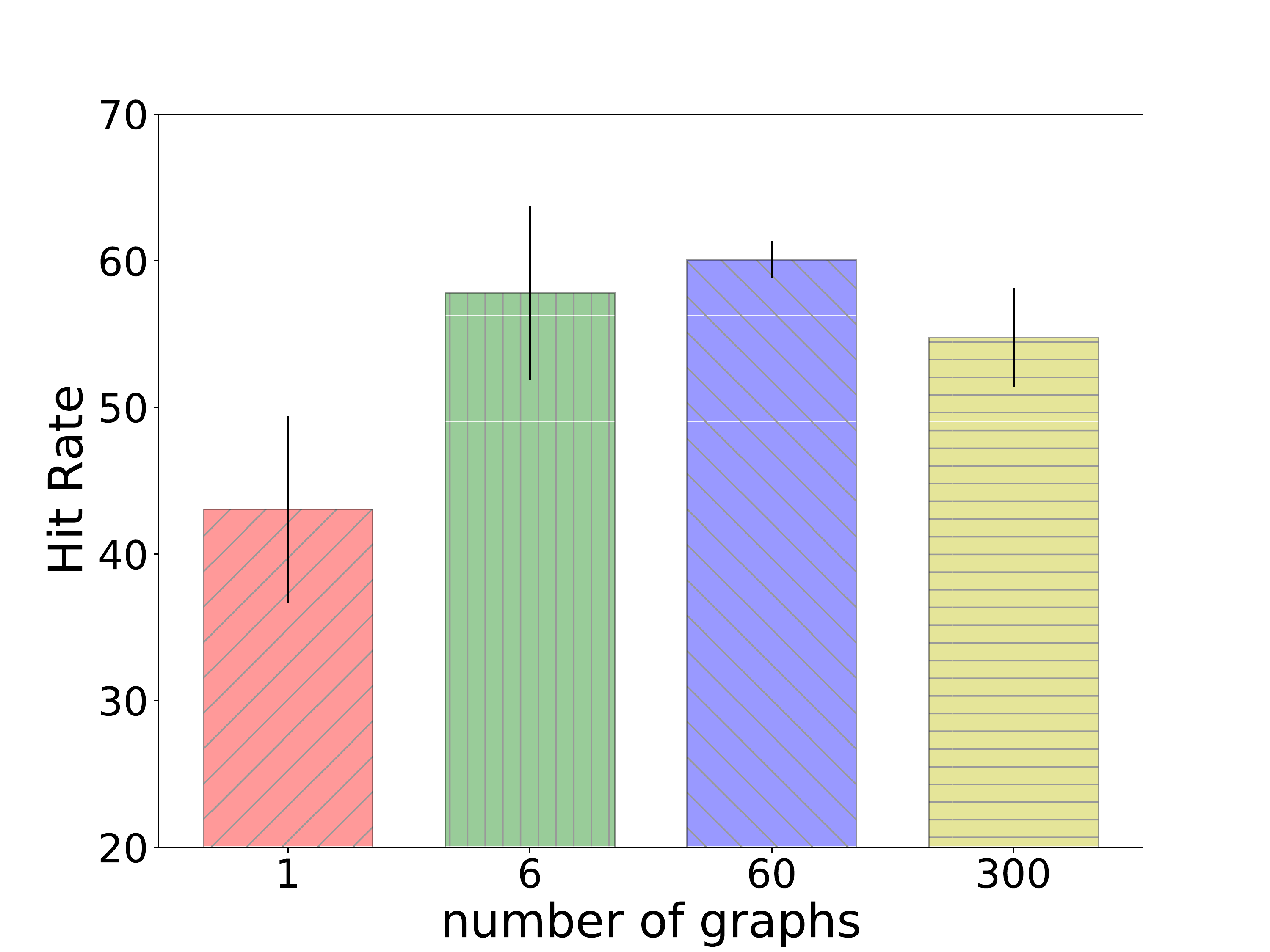}}}
\subfloat[Tianchi-61626 \label{fig:61-num}]{{\includegraphics[width=0.4\linewidth]{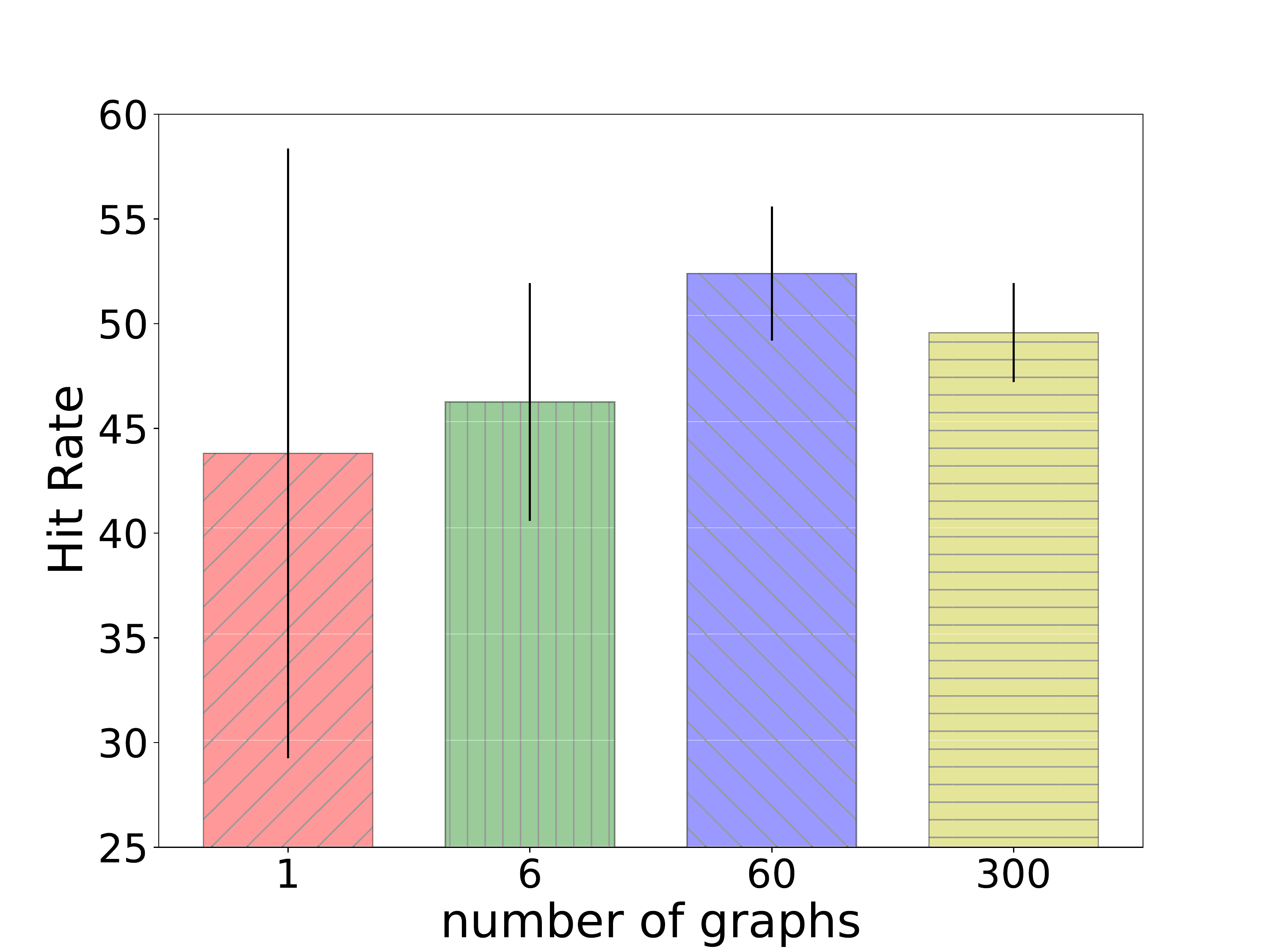} }}%

 \subfloat[Tianchi-393719 \label{fig:39-num}]{{\includegraphics[width=0.4\linewidth]{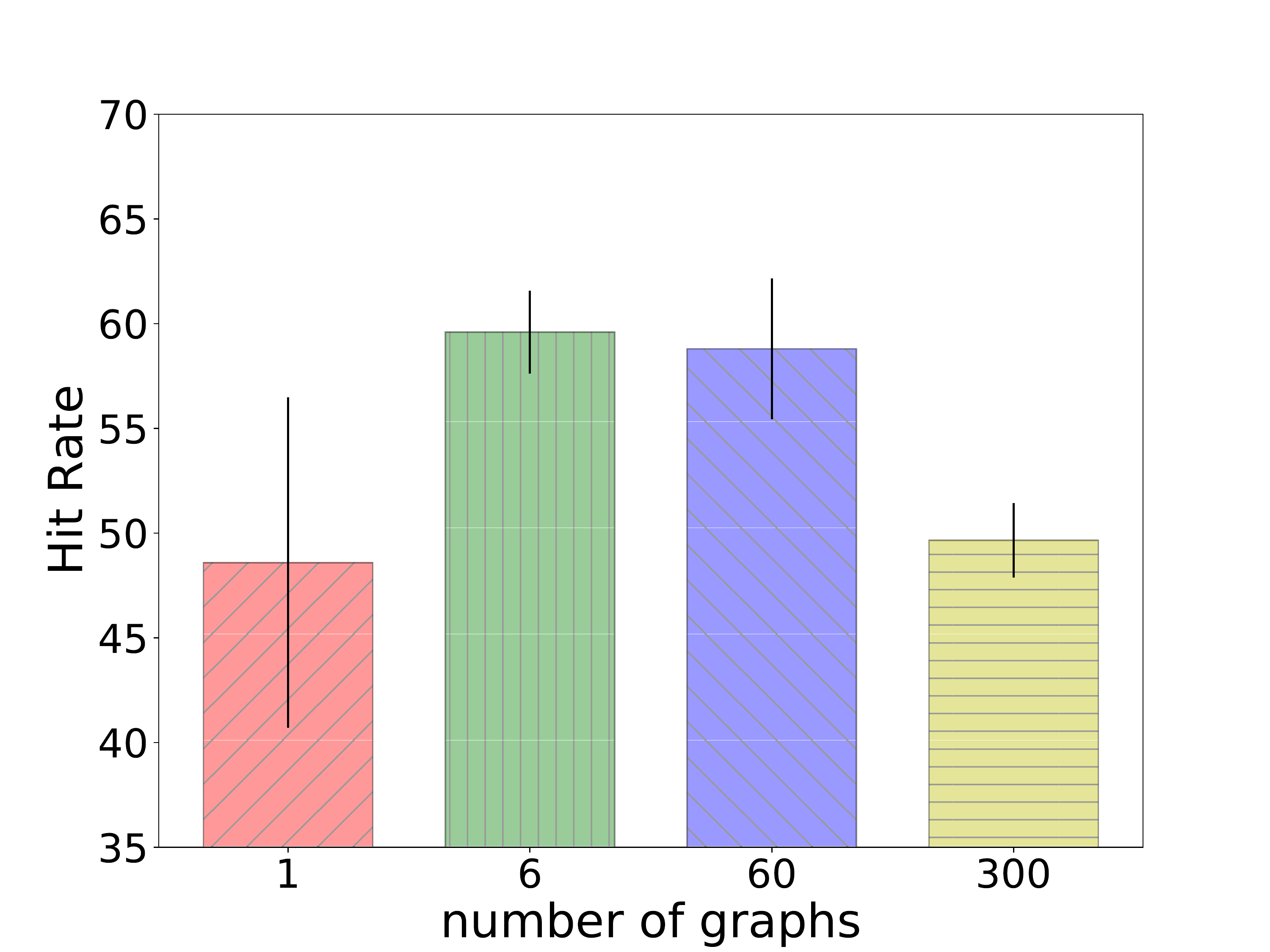}}}%
\subfloat[Tianchi-2798696 \label{fig:27-num}]{{\includegraphics[width=0.4\linewidth]{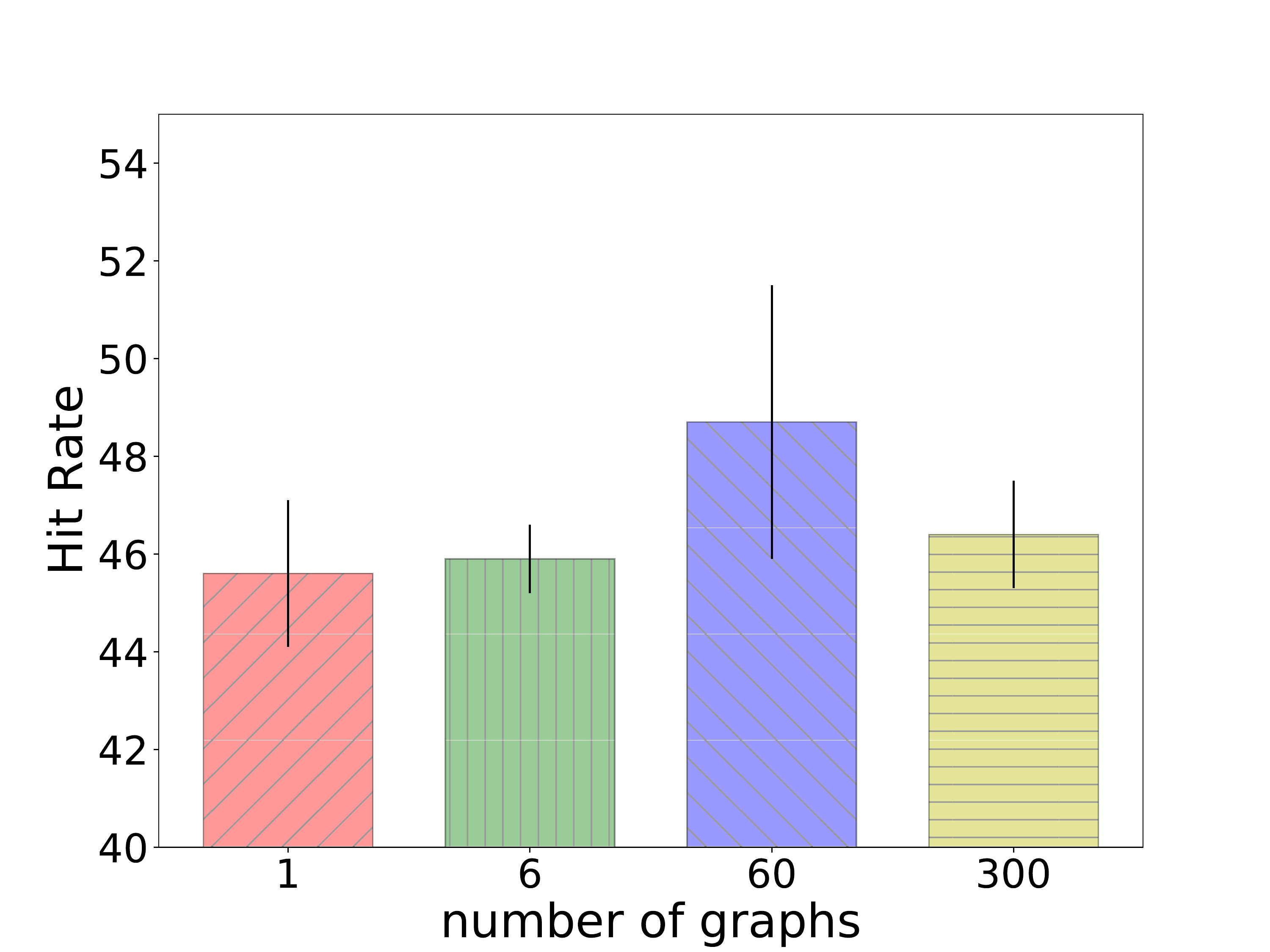} }}
 \caption{Performance variants of the proposed ADAPT in terms of the number of pre-training graphs.}%
\label{fig:num-graph}

\end{figure}
In order to explore the influence of the number of pre-training graphs on the model performance, we pre-train four ADAPT models on one single graph, 6 graphs, 60 graphs and 300 graphs, respectively. Then we fine-tune these models on four target downstream graphs. Note that for the ADAPT model pre-trained on multiple graphs, we report its best performance among two fine-tuning strategies. For the ADAPT model pre-trained on one single graph, it is not applicable to pre-train the GNN adaptor, thus we report its performance without the GNN adaptor. As shown in Figure~\ref{fig:num-graph}, the model performance first increases when the number of pre-training graphs increases and then it decreases consistently on all the datasets when the number of graphs is 300. This demonstrates that it is potentially beneficial to increase the number of pre-training graphs for the downstream performance. However, too many pre-training graphs may introduce noise that could hurt the downstream performance.

\section{Related Work}
Our work is related to graph neural networks, pre-training for graph neural networks and recommendation models based on graph Neural Network. Next, we briefly review representative methods from each category. 

\subsection{Graph Neural Networks}

In recent years, increasing attention and efforts have been devoted into graph neural networks, which successfully extend deep neural networks to graph data. Graph neural networks are theoretically and empirically demonstrated to be very powerful in graph representation learning~\cite{kipf2016semi,hamilton2017inductive,velivckovic2017graph}, and have achieved great success in various applications from different domains, such as natural language processing~\cite{yao2019graph,huang2019text,cai2020graph}, computer vision~\cite{qi2018learning,wang2019dynamic} and recommendation~\cite{yang2018hop,ying2018graph,wang2019neural,he2020lightgcn,DBLP:conf/www/WuCSHWW21,DBLP:conf/icde/JinZ00W20,DBLP:conf/icde/ZhengG00J20}. There are mainly two groups of graph neural networks: the spectral-based methods and the spatial-based methods.
In~\cite{scarselli2008graph}, the first graph neural network is proposed from the spatial perspective to solve both graph and node level tasks, which aggregates information from neighboring nodes for each node in every layer. Subsequently, Bruna et al.~\cite{bruna2013spectral} proposes to generalize the convolution operation to the graph domain based on graph Laplacian theory from the spectral perspective. Following this work, Defferrard et al~\cite{ChebNet} proposes ChebNet, which uses chebyshev polynomials to modulate the graph Fourier coefficients for different graph signals. Next, Kipf and Welling~\cite{kipf2016semi} further simplifies ChebNet via some assumptions and proposes graph convolutional networks (GCNs). It is remarkable that despite developed from the spectral perspective, GCNs can also be well illustrated in a spatial way. From then on, multiple spatial-based graph neural networks have been proposed. A comprehensive overview about GNNs can be found in recent surveys~\cite{wu2020comprehensive,zhang2020deep} and books~\cite{dlgraph}. In addition, there emerge some work trying to further explore the rationale behind GNNs, such as Ma et al~\cite{ma2020unified} proposes that most existent GNNs can be unified as graph signal denoising. In addition to the aforementioned work mainly focusing on graph convolution operation, there are also numerous works targeting at graph pooling operation, which summarizes graph representation from node representations and plays an essential role in graph representation learning. There are simple pooling methods, such as directly averaging all the node representations as the graph representation~\cite{duvenaud2015convolutional} or adding a virtual node that connects to all the nodes in the graph and then taking its node representation as the graph representation~\cite{li2015gated}. Besides, there are increasing number of hierarchical pooling methods proposed to learn graph representation hierarchically, which are believed to be able to better capture the graph structure information. Specifically, DiffPool~\cite{ying2018hierarchical} is proposed to learn a differentiable soft cluster assignment for each node at every GNN layer. Inspired by encoder-decoder model, graph U-Net~\cite{gao2019graph} is designed to consist of graph pooling (gPool) and graph unpooling (gUnpool) operations, where gPool can adaptively select important nodes based on the importance scores computed on a learnable projection vector. Furthermore, RepPool~\cite{li2020graph} proposes a learnable way to better integrate non-selected nodes, which can better preserve the information of both the important nodes and normal nodes .In addition, EigenPooling~\cite{ma2019graph} is designed from the perspective of graph Fourier transform and it can leverage both the node features and the local structures.

\subsection{Pre-training for Graph Neural Networks}
Inspired by the great success of pre-training techniques in multiple domains such as computer vision and natural language processing~\cite{mikolov2013efficient,radford2018improving,devlin2018bert,deng2009imagenet,he2019rethinking,huh2016makes}, many researchers have also started to explore how to a apply pre-training appropriately in GNN models~\cite{hu2019strategies,qiu2020gcc,hu2020gpt,lu2021learning,hao2021pre}. Hu et al. ~\cite{hu2019strategies} proposes a pre-training strategy based on self-supervised learning, which focuses on pre-training a GNN model at both the node level and graph level, so that the pre-trained model can learn effective node representations and graph representations simultaneously. Qiu et al. proposes GCC~\cite{qiu2020gcc}, a GNN pre-trainng framework based on contrastive learning, which aims at capturing transferable topological knowledge across multiple graphs. 

GPT-GNN~\cite{hu2020gpt} is a GNN pre-training framework built on generative model. It aims at capturing both the structure information and the semantic information via generating node attributes and edges alternatively. In addition, some work focus on bridging the gap between GNN pre-training and GNN fine-tuning. For example, L2P-GNN~\cite{lu2021learning} is proposed to mimic fine-tuning during the pre-training process via a dual-adaption mechanism at both the node level and graph level. Besides, there also emerges GNN pre-training work specially for alleviating cold-start problem in recommendation tasks~\cite{hao2021pre}, where researchers proposes to use embedding reconstruction as pre-training task, so that the cold-start user/item can get a good representation from the reconstruction knowledge.

\subsection{Graph Neural Networks for Recommendation Systems}

Recommendation tasks are to predict a user's preference based on its historical interactions with various items. Historical user-item interactions can be directly denoted as a bipartite graph, and thus it is very natural to apply graph neural networks to recommendation tasks. In fact, there exist numerous works focusing on exploring GNNs for recommendation systems and many of them have achieved promising performance~\cite{pinsage,wang2019neural,he2020lightgcn,fan2019graph,wang2019kgat}. PinSage~\cite{pinsage} is designed especially for recommendation tasks based on GraphSage, and can be directly applied to a web-scale recommendation tasks. For each user or item node, it utilizes a random-walk based sampling method to sample some neighboring nodes for information aggregation. NGCF~\cite{wang2019neural} is designed to explicitly capture high-order connectivity in user-item interaction graphs via embedding propagation. Later on, He et al. proposes LightGCN~\cite{he2020lightgcn}, which simplifies the design of GNNs especially for recommendation tasks via eliminating feature transformation and non-linear activation function, and has achieved significant performance improvement. Furthermore, Wu et al.~\cite{wu2021self} explored to improve the performance of GNN models for recommendations by incorporating self-supervised learning techniques. This work can be supplementary to most supervised GNN models for recommendations, which aim at improving user and item representations via self-discrimination.
These GNN-based recommendation methods only relying on the user-item interaction graphs. There are also works focusing on utilizing GNNs to deal with side information, such as social networks and knowledge graphs, to facilitate recommendation performance. For examples, GraphRec~\cite{fan2019graph} is proposed to use two graph attention networks to learn user embeddings and item embeddings, and the user embedding is learned from both the social graph and interaction graph. KGAT~\cite{wang2019kgat} is proposed to integrate the user-item interaction graph and the knowledge graph into one unified graph by viewing the user-item interaction as one type of the relations, and to use attentive embedding propagation layers to refine embeddings over this unified graph. 

\section{Conclusion}
In this paper, we propose an adaptive graph pre-training framework for localized collaborative filtering, ADAPT, which can effectively help alleviate the data scarcity challenge in recommendation tasks. There are two key components in the proposed ADAPT: the meta-LGNN and the GNN adaptor. Specifically, the meta-LGNN is a novel GNN-based recommendation method from a new perspective of local structure. It aims at encoding the collaborative filtering information into the graph representation for the neighboring structure of a given user-item pair, and it does not require learning user/item embeddings. Thus it is more flexible for pre-training compared to existing GNN-based recommendation methods. The GNN adaptor is the key to allow the effectiveness of taking advantage of multiple graphs for pre-training. It can capture the difference for each graph, and adapt the meta-LGNN model to the customized GNN model accordingly. Overall, the proposed ADAPT is able to transfer common knowledge from the pre-training graphs for the target downstream recommendation graph, and to capture their uniqueness in the meanwhile. Extensive experiments have validated the effectiveness and superiority of the proposed ADAPT in GNN pre-training for recommendations. 

There are a few directions we can further explore in the future as follows. First, the current implementation of the meta-LGN is based on GCN, which is a vanilla model in graph representation learning. The performance of the meta-LGN could be further improved by replacing the GCN model with other advanced GNN models such as GIN. Second, the node labeling method currently adopted by the meta-LGN treats user and item nodes equally. In the further exploration, we may consider to distinguish the user and item nodes in the node labeling method. Third, the designs of the adaptor model and adaptation operation are also flexible, and there are some alternative models and operations that can further be explored on the proposed framework. Last but not least, instead of pre-training the proposed framework on numerous totally independent user-item interaction graphs, we may consider how to leverage some available shared user or item nodes among theses graphs.

\newpage
\bibliographystyle{ACM-Reference-Format}
\bibliography{sample-base}

\end{document}